\documentclass[fleqn,usenatbib]{mnras}

\usepackage{newtxtext,newtxmath}

\usepackage[T1]{fontenc}
\usepackage{ae,aecompl}

\usepackage{array}
\newcolumntype{?}{!{\vrule width 1.5pt}}
\usepackage{afterpage}

\usepackage{multirow}
\usepackage{graphicx}	
\usepackage{amsmath}	
\usepackage{amssymb}	
\usepackage[usenames]{color}
\usepackage{bm}		
\usepackage{booktabs,makecell}
\usepackage{adjustbox}
\usepackage{hyperref}
\usepackage{soul}
\usepackage[switch,pagewise]{lineno}

\usepackage{eso-pic}

\AddToShipoutPictureBG*{%
  \AtPageUpperLeft{%
    \hspace{0.75\paperwidth}%
    \raisebox{-3.5\baselineskip}{%
      \makebox[0pt][l]{\textnormal{DES 2017-0290}}
}}}%

\AddToShipoutPictureBG*{%
  \AtPageUpperLeft{%
    \hspace{0.75\paperwidth}%
    \raisebox{-4.5\baselineskip}{%
      \makebox[0pt][l]{\textnormal{FERMILAB-PUB-17-594-AE}}
}}}%

\makeatletter
\def\blfootnote{\xdef\@thefnmark{}\@footnotetext}
\makeatother

\title[DES Y1: Calibration of redMaGiC Redshift Distributions]{Dark Energy Survey Year 1 Results: Calibration of redMaGiC Redshift Distributions in DES and SDSS from Cross-Correlations}

\author[R. Cawthon et al.]{
\parbox{\textwidth}{
\Large
R.~Cawthon$^{1,2*}$,
C.~Davis$^{3}$,
M.~Gatti$^{4}$,
P.~Vielzeuf$^{4}$,
J.~Elvin-Poole$^{5}$,
E.~Rozo$^{6}$,
J.~Frieman$^{7,1,2}$,
E.~S.~Rykoff$^{3,8}$,
A.~Alarcon$^{9}$,
G.~M.~Bernstein$^{10}$,
C.~Bonnett$^{4}$,
A.~Carnero~Rosell$^{11,12}$,
F.~J.~Castander$^{9}$,
C.~Chang$^{1}$,
L.~N.~da Costa$^{11,12}$,
J.~De~Vicente$^{13}$,
J.~DeRose$^{14,3}$,
A.~Drlica-Wagner$^{7}$,
E.~Gaztanaga$^{9}$,
T.~Giannantonio$^{15,16,17}$,
D.~Gruen$^{3,8}$,
J.~Gschwend$^{11,12}$,
W.~G.~Hartley$^{18,19}$,
B.~Hoyle$^{20,17}$,
H.~Lin$^{7}$,
M.~A.~G.~Maia$^{11,12}$,
R.~Miquel$^{21,4}$,
R.~L.~C.~Ogando$^{11,12}$,
M.~M.~Rau$^{17}$,
A.~Roodman$^{3,8}$,
A.~J.~Ross$^{22}$,
I.~Sevilla-Noarbe$^{13}$,
M.~A.~Troxel$^{22,23}$,
R.~H.~Wechsler$^{14,3,8}$,
T.~M.~C.~Abbott$^{24}$,
F.~B.~Abdalla$^{18,25}$,
S.~Allam$^{7}$,
J.~Annis$^{7}$,
S.~Avila$^{26,27}$,
M.~Banerji$^{15,16}$,
K.~Bechtol$^{28}$,
R.~A.~Bernstein$^{29}$,
E.~Bertin$^{30,31}$,
D.~Brooks$^{18}$,
D.~L.~Burke$^{3,8}$,
M.~Carrasco~Kind$^{32,33}$,
J.~Carretero$^{4}$,
C.~E.~Cunha$^{3}$,
C.~B.~D'Andrea$^{10}$,
D.~L.~DePoy$^{34}$,
S.~Desai$^{35}$,
H.~T.~Diehl$^{7}$,
P.~Doel$^{18}$,
T.~F.~Eifler$^{36,37}$,
A.~E.~Evrard$^{38,39}$,
B.~Flaugher$^{7}$,
P.~Fosalba$^{40}$,
J.~Garc\'ia-Bellido$^{27}$,
D.~W.~Gerdes$^{38,39}$,
R.~A.~Gruendl$^{32,33}$,
G.~Gutierrez$^{7}$,
D.~Hollowood$^{41}$,
K.~Honscheid$^{22,23}$,
D.~J.~James$^{42}$,
T.~Jeltema$^{41}$,
E.~Krause$^{36,37}$,
K.~Kuehn$^{43}$,
S.~Kuhlmann$^{44}$,
N.~Kuropatkin$^{7}$,
O.~Lahav$^{18}$,
M.~Lima$^{45,11}$,
J.~L.~Marshall$^{34}$,
P.~Martini$^{22,46}$,
F.~Menanteau$^{32,33}$,
C.~J.~Miller$^{38,39}$,
A.~A.~Plazas$^{37}$,
E.~Sanchez$^{13}$,
V.~Scarpine$^{7}$,
R.~Schindler$^{8}$,
M.~Schubnell$^{39}$,
E.~Sheldon$^{47}$,
M.~Smith$^{48}$,
R.~C.~Smith$^{24}$,
M.~Soares-Santos$^{7}$,
F.~Sobreira$^{49,11}$,
E.~Suchyta$^{50}$,
M.~E.~C.~Swanson$^{33}$,
G.~Tarle$^{39}$,
D.~Thomas$^{26}$,
D.~L.~Tucker$^{7}$,
A.~R.~Walker$^{24}$
\begin{center} (DES Collaboration) \end{center}
}
\vspace{0.4cm}
\\
}

\date{Accepted XXX. Received YYY; in original form ZZZ}

\pubyear{2017}

\begin{document}
\label{firstpage}
\pagerange{\pageref{firstpage}--\pageref{lastpage}}
\maketitle

\begin{abstract}
We present calibrations of the redshift distributions of redMaGiC galaxies in the Dark Energy Survey Year 1 (DES Y1) and Sloan Digital Sky Survey (SDSS) DR8 data. These results determine the priors of the redshift distribution of redMaGiC galaxies, which were used for galaxy clustering measurements and as lenses for galaxy-galaxy lensing measurements in DES Y1 cosmological analyses. We empirically determine the bias in redMaGiC photometric redshift estimates using angular cross-correlations with Baryon Oscillation Spectroscopic Survey (BOSS) galaxies.  For DES, we calibrate a single parameter redshift bias in three photometric redshift bins: $z \in[0.15,0.3]$, [0.3,0.45], and [0.45,0.6]. Our best fit results in each bin give photometric redshift biases of $|\Delta z|<0.01$. To further test the redMaGiC algorithm, we apply our calibration procedure to SDSS redMaGiC galaxies, where the statistical precision of the cross-correlation measurement is much higher due to a greater overlap with BOSS galaxies. For SDSS, we also find best fit results of $|\Delta z|<0.01$. We compare our results to other analyses of redMaGiC photometric redshifts.
\end{abstract}

\begin{keywords}
galaxies: distances and redshifts -- large-scale structure of Universe -- surveys
\end{keywords}
\blfootnote{Affiliations are listed at the end of the paper.}
\blfootnote{*e-mail: rcawthon@oddjob.uchicago.edu}
\setcounter{footnote}{1}



\section{Introduction}

The Dark Energy Survey is an on-going, five-year photometric survey, which will image $5,000 \ \text{deg}^2$ of the sky. In \cite{keypaper}, a cosmological analysis is produced based on measurements of DES year 1 data including galaxy clustering \citep{wthetapaper}, cosmic shear \citep{shearcorr} and galaxy-galaxy lensing \citep{gglpaper}. The cosmological interpretation of these measurements relies critically on precise and accurate estimates of galaxy redshift distributions. Redshifts at cosmological distances indicate a specific time in the universe being observed, making them paramount for each of these measurements to accurately study the history of the universe. While spectroscopic surveys can obtain precise galaxy redshifts, currently they cannot sample the large sky areas to faint enough magnitudes needed for the above cosmological measurements. As a result, we must rely on multi-band photometric surveys to provide approximate redshift estimates.

Photometric galaxy surveys such as DES \citep{DES}, KiDS \citep{KIDS}, HSC \citep{HSC} and in the future LSST \citep{LSST}, Euclid \citep{EUCLID}, and WFIRST \citep{WFIRST} will rely on this technique of estimating approximate photometric redshift (photo-z) estimates from multi-band imaging. Reviews of photo-z algorithms can be found in e.g., \cite{Hildebrandt2010}, \cite{Sanchez2014}, \cite{bonnett2016} and \cite{photoz} and references therein. The robustness of their cosmological results will depend upon the reliability of the photo-z estimates.

In the past decade, a separate technique of determining redshifts has been studied, starting most prominently with \cite{schneider2006} and \cite{newman2008}. The technique, sometimes called the cross-correlation method, or clustering redshifts, involves measuring angular correlation functions between a sample of galaxies for which the redshifts are unknown, and a reference sample of galaxies for which the redshifts are known. The technique uses the fact that because galaxies cluster via gravity, two galaxies with small angular separation are more likely to be spatially correlated and thus at similar distances (and redshifts). The above is only a statistical statement since a pair of galaxies may be close on the sky due to chance projection. However, for a large sample of galaxies with unknown redshifts, angular clustering with a reference sample of known redshifts provides an informative prior on the redshift distribution of the former. The technique has developed with several variations in the past decade, and has been tested on multiple datasets and simulations (\cite{MattewsNewman2010}, \cite{Menard2013}, \cite{Schmidt2013}, \cite{McQuinn2013} among others). Recent uses include calibrating redshifts of CFHTLenS \citep{cfhtlens}  galaxies in \cite{Choi2016}, of KiDS galaxies in \cite{Hildebrandt2017}, \cite{Morrison:2017aa} and \cite{Johnson2017}, and DES science verification galaxies in \cite{DAVIS}. 

This work focuses on calibrating the redshift distribution of redMaGiC\footnote{The name redMaGiC stands for red sequence Matched-filter Galaxy Catalog} galaxies using cross-correlations. These galaxies are luminous red galaxies (LRG) selected by the redMaGiC algorithm \citep{redmagicSV}. The algorithm was specifically designed to create a sample of LRGs with minimal photometric redshift errors. The resulting sample is also luminosity-thresholded and has a constant comoving density. Each of these features is important for measurements of large-scale structure.  The redMaGiC algorithm relies on the redMaPPer red sequence cluster finder \citep{rykoff+14}, which uses a set of spectroscopic galaxy clusters and photometric data to create a photometric template for the red sequence of galaxies as a function of redshift. The redMaGiC algorithm selects galaxies when its colors are well matched by the template.

Several DES Year 1 studies use redMaGiC galaxies. In \cite{wthetapaper}, the spatial clustering of redMaGiC galaxies is measured. In \cite{gglpaper}, redMaGiC galaxies are used as the lenses in galaxy-galaxy lensing measurements. Both of these measurements are used in the cosmological analysis of \cite{keypaper}. Separately, redMaGiC galaxies are used for a counts and lensing in cells cosmological analysis in \cite{troughs}. Uncertainties and biases in the redshift distributions of the redMaGiC galaxies will contribute to statistical and systematic errors in the cosmological parameter estimates derived from these measurements. 

Two other cross-correlations papers using DES Y1 data, \cite{xcorr} and \cite{xcorrtechnique}, also use the redMaGiC galaxies. In these analyses, the redMaGiC galaxies serve as a reference sample to cross-correlate with the weak lensing source galaxies \citep{shearcat}, with \cite{xcorr} calibrating the Y1 data and \cite{xcorrtechnique} using simulations to assess the systematic uncertainties of the method. The redMaGiC photo-z's are much more precise than those for the weak lensing source galaxies. The redMaGiC galaxies can thus be used as a ``pseudo-spectroscopic" sample to calibrate the redshift distribution of the source galaxies via cross-correlations. Using redMaGiC as the reference sample is necessary since there are too few galaxies with spectroscopic redshifts in the DES footprint to use for cross-correlations to calibrate the source galaxies. This paper's calibration of redMaGiC thus also impacts these other cross-correlation papers and their constraints used in the DES Y1 cosmology papers.

To calibrate redMaGiC, we cross-correlate with the LOWZ and CMASS spectroscopic galaxy samples from the Baryon Oscillation Spectroscopic Survey (BOSS, \cite{2013AJ....145...10D} from the Sloan Digital Sky Survey \citep{2000AJ....120.1579Y} Data Release 12 (SDSS DR12, \cite{SDSSdr12}). The BOSS sample was chosen since it is the largest spectroscopic sample that overlaps the DES Y1 footprint. The overlap overall is a small fraction of the full DES Y1 footprint though, only over part of the region known as Stripe 82 (RA $\in [317,360]$, DEC $\in [-1.8,1.8]$). Each of the DES analyses using redMaGiC mentioned previously do not use this Stripe 82 area of Y1, opting only to use the much larger contiguous region further South. The Stripe 82 DES redMaGiC sample is roughly $10 \%$ as large as the Southern redMaGiC sample used in the other analyses. The Stripe 82 and Southern DES redMaGiC samples were created with the same methodology.

The main goal of this paper is to estimate the photo-z bias of the redMaGiC algorithm to support the DES Y1 cosmological measurements. A secondary goal is to understand the redMaGiC photo-z biases in more detail to support future uses of the algorithm. For this second goal, we also study the SDSS redMaGiC sample which has a far larger overlap with the BOSS spectroscopic galaxies, allowing us to greatly reduce the statistical errors on our calibrations. In the limit that the DES and SDSS redMaGiC photo-z's behave similarly, the SDSS results may be the more precise measurement of issues also present in DES. In Section \ref{sec:photozerrors}, we compare the measured photo-z biases with the estimated biases of the redMaGiC algorithm in \cite{redmagicSV}. A large sample of SDSS redMaGiC galaxies that have spectroscopic redshifts (spec-z) will also be used to test our methodology and systematics in Section \ref{sec:sdss_tests}.

The outline of this paper is as follows. In Section \ref{sec:datasets}, we describe the datasets used in our work, redMaGiC galaxies and reference spectroscopic galaxies. In Section \ref{sec:methods}, we describe our methodology of measuring the cross-correlations, correcting for galaxy clustering bias evolution and using our results to calculate a photometric redshift bias for redMaGiC. (From this point forward, `galaxy bias' will always refer to the linear galaxy clustering bias, $b$ in Equation \ref{crosscorr}, not to be confused with the photo-z bias.) In Section \ref{sec:systematics}, we validate our methodology by testing on a subset of SDSS redMaGiC galaxies that have spectroscopic redshifts, and estimate the amplitude of different systematic uncertainties (primarily the galaxy bias evolution) based on that test and others with the main datasets. In Section \ref{sec:results}, we present our main results for DES and SDSS redMaGiC photo-z biases in our fiducial redshift binning. In Section \ref{sec:photozerrors}, we alter our analysis somewhat to more precisely estimate redMaGiC photo-z biases as a function of redshift and compare to previous estimates of bias. In Section \ref{sec:summary}, we summarize our work. 

\section{Datasets}
\label{sec:datasets}

\subsection{Dark Energy Survey redMaGiC}
\label{sec:des}

The redMaGiC selection algorithm of LRGs is described in \cite{redmagicSV}.  The algorithm has been slightly updated as described in \cite{wthetapaper}, and applied to the DES Y1 Gold catalog \citep{y1gold}. For DES Y1, the redMaGiC algorithm uses a sampling from a Gaussian distribution ($z_{\text{rmg}} \pm \sigma_{\text{rmg}}$) rather than simply the central redshift value, $z_{\text{rmg}}$, to compute the comoving density as done in \cite{redmagicSV}.

We calibrate the photometric redshift bias of redMaGiC galaxies in the photometric redshift bins defined in \cite{wthetapaper}, namely $z \in [0.15,0.3],[0.3,0.45],[0.45,0.6]$. We note that while \cite{wthetapaper} defines two additional higher-redshift bins, the number of spectroscopic galaxies at these redshifts is too low for us to use cross-correlation techniques for photo-z calibration. 

As described in \cite{wthetapaper}, there are three redMaGiC samples defined by luminosity cuts of $L/L_{*} >0.5$ (`high-density'), $L/L_{*}>1.0$ (`high-luminosity') and $L/L_{*}>1.5$ (`higher-luminosity') where the reference luminosity $L_{*}$ is computed using a Bruzual and Charlot \citep{bruzualcharlot03} model for a single star-formation burst at $z=3$ \citep{Rykoff:2016aa}. The samples use different photometric methods for the red sequence training. The high-density sample uses SExtractor MAGAUTO quantities applied to redMaPPer as in \cite{soergel+16}. The training for the high-luminosity and higher-luminosity samples uses a multi-epoch, multi-band and multi-object fit (MOF) described in \cite{y1gold} applied to redMaPPer in \cite{Mcclintock17}. We also use a DES systematics derived set of weights as used in \cite{wthetapaper}. More details on the redMaGiC catalogs are in \cite{wthetapaper}.

For the redshift range of $z=0.15-0.6$, the high-density sample is what is used by \cite{wthetapaper} and \cite{keypaper}. Those papers use the high- and higher-luminosity sample only at $z>0.6$, a redshift range that our work cannot study due to fewer spectroscopic galaxies in BOSS. However, the higher-luminosity sample at $0.15<z<0.6$ is used in \cite{xcorr}, as a reference sample for cross-correlations with weak lensing source galaxies used in \cite{shearcorr}, \cite{gglpaper} and \cite{keypaper}. Since no DES Y1 analysis uses the high-luminosity sample in the redshift range we probe, we only calibrate the high-density and higher-luminosity samples for DES.

The redMaGiC algorithm assigns each galaxy in the catalog a redshift value, $z_{\text{rmg}}$ and error $\sigma_{\text{rmg}}$. The redshift value, $z_{\text{rmg}}$, is used to place the galaxies into the different redshift bins. However, the photometric redshift distributions, $n_{\text{pz}}$ are built by stacking estimates of the redshift of each galaxy assuming the probability distribution function (PDF) is a Gaussian centered at $z_{\text{rmg}}$ with spread $\sigma_{\text{rmg}}$. Thus, the photometric redshift distribution of e.g., a bin of $z \in [0.15,0.3]$ will extend into $z<0.15$ and $z>0.3$. All plotted photometric redshift distributions in this work match this procedure, which is also done in \cite{wthetapaper}.

The redMaGiC sample in Stripe 82 (RA $\in [317,360]$, DEC $\in [-1.8,1.8]$), where we have the ability to use cross-correlations with BOSS galaxies, spans $\sim 124 \ \text{deg}^2$ after masking. This sample is about $10\%$ the size of the main DES redMaGiC sample, which is further South (roughly RA $\in [300,360],[0,100]$, DEC $\in [-60,-40]$, see \cite{wthetapaper}). Numbers for the DES redMaGiC samples in this region used in this work are given in Table \ref{table:des_ngals}.

The spatial separation between the redMaGiC galaxies used in this work for calibration and those in the main DES cosmology papers theoretically could mean the calibrations are not applicable to the DES Y1 papers. Procedures and tests in \cite{y1gold} and \cite{wthetapaper} strongly limit the extent these samples could be different though. The Gold catalog created in \cite{y1gold} contains all of the area covered by this work (`Stripe 82 region') and the cosmology papers (`SPT region'). \cite{y1gold} constrains the photometric calibration to within 2\% across the full footprint. That calibration includes accounting for Galactic dust reddening as measured by \cite{SFD} by using stellar locus regression \citep{kelly2014}. \cite{wthetapaper} goes a step further in assessing systematics for the redMaGiC catalog by looking for correlations of galaxy density with various systematics, such as seeing, exposure time, stellar density, Galactic extinction and other properties to create a system of weights for the redMaGiC catalog. This same process was done for both the `Stripe 82' and `SPT' redMaGiC catalogs. The resulting weights had negligible impact on the cross-correlations measurements in this paper. This is not surprising, as the systematics in \cite{wthetapaper} were seen to be significant only at large scales ($>60'$). This work uses smaller scales than all of the measurements in \cite{wthetapaper}, with their work using only scales $>10'$, and ours using only scales $<10'$, after converting from the physical distance bounds mentioned in Section \ref{sec:methods}. Furthermore, our methodology weights the smallest angular scales (Equation \ref{DavisPeeblesestimator}).

\begin{table}
\begin{center}
    \begin{adjustbox}{width=0.5\textwidth}
    \begin{tabular}{|c|c|c|c|}
      \hline
      DES Galaxy Sample & $L/L_{*}$ & $n_{\text{gal}}(\text{arcmin}^{-2})$ & $N_{\text{gal}}$ \\
      \hline
      High-density ($z \in [0.15,0.3]$) & 0.5 & 0.0149 & 6660  \\   
      \hline
      High-density ($z \in [0.3,0.45]$) & 0.5 & 0.0335 & 14952  \\   
      \hline
      High-density ($z \in [0.45,0.6]$) & 0.5 & 0.0529 & 23634  \\   
      \hline
      Higher-luminosity ($z \in [0.15,0.3]$) & 1.5 & 0.0020 & 912  \\   
      \hline
      Higher-luminosity ($z \in [0.3,0.45]$) & 1.5 & 0.0039 & 1731  \\   
      \hline
      Higher-luminosity ($z \in [0.45,0.6]$) & 1.5 & 0.0069 & 3089  \\   
      \hline
    \end{tabular}
  \end{adjustbox}
  \caption{DES redMaGiC number of galaxies by sample used in this work. The high-density and higher-luminosity samples are defined by the luminosity threshold, $L/L_{*}$. The total number of galaxies, $N_{\text{gal}}$ and the galaxy density, $n_{\text{gal}}$ are also shown. These samples span $\sim 124 \ \text{deg}^2$, and are approximately $10\%$ the size of the separate main DES Year 1 sample described in Elvin-Poole et al. 2017.}
  \label{table:des_ngals}
  \end{center}
\end{table}

\subsection{Sloan Digital Sky Survey redMaGiC (DR8)}
\label{sec:sdss}

We use the SDSS DR8 redMaGiC catalogs created in \cite{redmagicSV}. SDSS DR8 photometric data is described in \cite{2011ApJS..193...29A}. The catalogs use the red sequence calibration of the DR8 redMaPPer catalog \citep{rykoff+14}. Masking of the DR8 galaxy catalog was done while applying the redMaPPer and redMaGiC algorithms, using data from the mask in the Baryon Acoustic Oscillation Survey (BOSS) \citep{2013AJ....145...10D}, as well as stellar masking using data in the Yale Bright Star Catalog \citep{1991bsc..book.....H} and New General Catalog (NGC, \cite{sinnott}). Spectroscopic training for the redMaPPer algorithm used SDSS DR10 data \citep{2014ApJS..211...17A}. The final catalog covers $\sim 9,350 \ \text{deg}^2$. 

Similar to the DES redMaGiC catalogs, the SDSS catalogs are defined by luminosity cuts. In \cite{redmagicSV}, the $L/L_*>0.5$ and $L/L_*>1.0$ catalogs are called `Faint' and `Bright' respectively, but we will refer to them by their DES equivalent names, `high-density' and `high-luminosity'. There is no `higher-luminosity' ($L/L_*$) equivalent catalog in the SDSS data, so we will calibrate the two available samples. The SDSS redMaGiC samples do not reach as large redshifts as DES does. We only analyze SDSS redMaGiC in our first two redshift bins, $z \in [0.15,0.3]$ and $z \in [0.3,0.45]$.

Particularly noteworthy for our work is a subset of the SDSS redMaGiC galaxies that have spectroscopic redshifts. For the high-density catalog, this includes $8.8\%$ of the Bin 1 ($z \in [0.15,0.3]$) galaxies, and $5.0\%$ of the Bin 2 ($z \in [0.3,0.45]$) galaxies. For the SDSS high-luminosity catalog, the two bins have spec-z measurements for $24.7\%$ and $12.2\%$ of the galaxies respectively. In Section \ref{sec:sdss_tests}, we test our methodology on this sample of galaxies where we are able to compare our estimates of the redshift distribution from clustering with the true redshift distribution given by spectroscopic redshifts. Since the high-density subsample with spec-z contains all of the high-luminosity galaxies with spec-z, and less than $25\%$ additional galaxies, we just analyze the high-density with spec-z subsample in Section \ref{sec:sdss_tests}. The high-luminosity with spec-z sample yields similar qualitative results as shown in that section. 

The photometric redshift distributions for SDSS redMaGiC are again built using Gaussian PDFs with mean $z_{\text{rmg}}$ and spread $\sigma_{\text{rmg}}$ as described for the DES redMaGiC galaxies. The total numbers of galaxies used in each of our SDSS redMaGiC datasets are shown in Table \ref{table:sdss_ngals}.

\begin{table}
\begin{center}
    \begin{adjustbox}{width=0.5\textwidth}
    \begin{tabular}{|c|c|c|c|}
      \hline
      SDSS Galaxy Sample & $L/L_{*}$ & $n_{\text{gal}} (\text{arcmin}^{-2})$ & $N_{\text{gal}}$ \\
      \hline
      High-density w/spec-z ($z \in [0.15,0.3]$) & 0.5 & 0.0013 & 43181  \\   
      \hline
      High-density w/spec-z ($z \in [0.3,0.45]$) & 0.5 & 0.0016 & 55214  \\   
      \hline
      High-density ($z \in [0.15,0.3]$) & 0.5 & 0.0152 & 512380  \\   
      \hline
      High-density ($z \in [0.3,0.45]$) & 0.5 & 0.0365 & 1228418  \\   
      \hline
      High-luminosity ($z \in [0.15,0.3]$) & 1.0 & 0.0031 & 102753  \\   
      \hline
      High-luminosity ($z \in [0.3,0.45]$) & 1.0 & 0.0074 & 247406  \\   
      \hline
    \end{tabular}
  \end{adjustbox}
  \caption{SDSS redMaGiC number of galaxies by sample. The subsample with spec-z will be used as a test sample in Section \ref{sec:sdss_tests}. The high-density and high-luminosity samples are defined by the luminosity threshold, $L/L_{*}$. The total number of galaxies, $N_{\text{gal}}$ and the galaxy density, $n_{\text{gal}}$ are also shown. This samples covers roughly the entire SDSS footprint, $\sim 9,350\ \text{deg}^2$.}
  \label{table:sdss_ngals}
  \end{center}
\end{table}

\subsection{Baryon Oscillation Spectroscopic Survey Galaxies (SDSS DR12)}
\label{sec:boss}

The spectroscopic sample used in our cross-correlations is the large-scale structure catalog described in \cite{2016MNRAS.455.1553R} from the BOSS survey \citep{2013AJ....145...10D} published as part of the SDSS DR12. The catalog, associated mask and simulated random galaxies are described in detail in \cite{2016MNRAS.455.1553R} and references therein. The random galaxy catalogs (used in Equation \ref{DavisPeeblesestimator}) correct issues in clustering measurements that can be created due to the masked areas of the sky, and edge-effects of the dataset. The random catalogs have far more galaxies than the data in order to not add Poisson noise. The galaxy catalog is comprised of two distinct samples known as LOWZ and CMASS. As done in \cite{2016MNRAS.455.1553R}, we split the two samples at $z=0.43$ to avoid overlap, removing the LOWZ galaxies above and the CMASS galaxies below that point. Using only one sample at each redshift alleviates concerns of how to properly combine galaxy catalogs, masks and random catalogs, and how this may affect the galaxy bias evolution which plays a large role in our methodology (Section \ref{sec:32}). We did not test using both the LOWZ and CMASS samples at overlapping redshifts (i.e. not applying the 0.43 cut), but doing so would only increase the sample of galaxies by $\sim 6\%$. The DR12 catalog was designed primarily to measure the baryon acoustic oscillations (BAO) signal \citep{AlamBOSSDR12}, but its properties as a large, uniformly selected spectroscopic galaxy sample fit the purposes of cross-correlations quite well.

\begin{figure}
\begin{center}
\includegraphics[width=0.5 \textwidth]{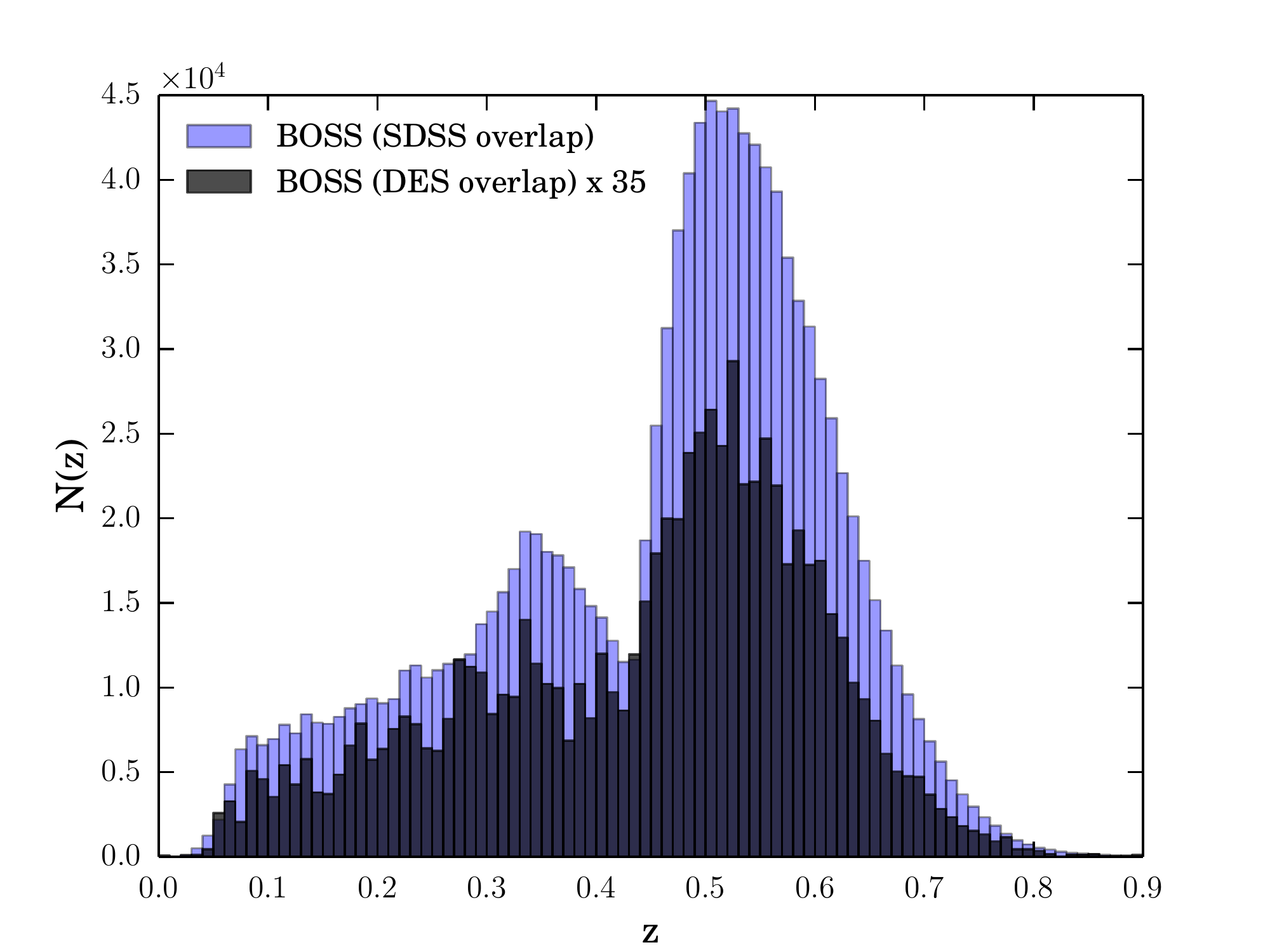}
\end{center}
\caption{The redshift distribution of the BOSS spectroscopic galaxies from SDSS DR12. The full sample that overlaps SDSS and the subsample that overlaps DES are both shown. The BOSS galaxies are our reference sample to cross-correlate with redMaGiC galaxies.}
\label{fig:bossdndz}
\end{figure}

For the correlations with DES redMaGiC, the BOSS catalog has 20,473 galaxies in Stripe 82 in the redshift range of $z \in [0.1,0.7]$. For the correlations with SDSS redMaGiC, we can use the full area of the catalog, which has 825,751 galaxies in the redshift range of $z \in [0.1,0.55]$. The full redshift distribution of the BOSS catalog is shown in Figure \ref{fig:bossdndz}, including redshift ranges not used in this work.

\subsection{Buzzard Simulations}
\label{sec:sims}

We make use of simulations in this work in Section \ref{sec:autocorrs}, where we use them to help characterize the the evolution in galaxy bias of the redMaGiC galaxies which can impact the cross-correlations (Section \ref{sec:32}). We use the Buzzard v1.1 simulation of the DES Y1 sample described in \cite{DeRose2017}, \cite{Wechsler2017}, and \cite{simspaper}. The galaxy catalogs are made from N-body simulations using L-Gadget 2, a modified version of GADGET2 \citep{springel2005} with haloes identified using ROCKSTAR \citep{Behroozi2013}, and galaxies assigned using ADDGALS \citep{Wechsler2017}. Spectral Energy Distributions (SEDs) are assigned using spectroscopic data from SDSS DR7 \citep{Cooper2011}. From these, photometry is generated for the DES filters and photometric errors are assigned using the Y1 depth map \citep{Rykoff15}. The redMaGiC algorithm is run on these photometric measurements.

We note that we do not test our full methodology in simulations due to uncertainties in modeling characteristics of the BOSS spectroscopic sample, such as its galaxy bias evolution. However, many of the choices for the cross-correlation methodology used in this work (Section 3.1) are based on testing cross-correlations of simulated DES redMaGiC galaxies and weak lensing source galaxies in \cite{xcorrtechnique}, which use the same Buzzard simulations as described here. For extensions beyond the methodology in \cite{xcorrtechnique}, such as testing the accuracy of the galaxy bias correction (Section \ref{sec:sdss_tests}), we use the SDSS redMaGiC galaxies with spectroscopic redshifts as a sample that we can validate our methods against.

\section{Methods}
\label{sec:methods}

\subsection{Unknown and Reference Correlation Measurement}
\label{sec:31}

The clustering redshifts method involves a cross-correlation between an `unknown' sample for which redshift estimates are desired, and a `reference' sample with known redshift measurements for each object. In our study (in contrast to \cite{xcorrtechnique} and \cite{xcorr}), the unknown is the redMaGiC galaxy sample, and the reference are the LOWZ and CMASS spectroscopic galaxies from BOSS/SDSS DR12.

The basic framework of the clustering measurement in this work is similar to that of \cite{xcorrtechnique}, and we provide a brief summary of the methods described there. Our analysis differs from that of \cite{xcorrtechnique} and \cite{xcorr} in our methodology to correct for galaxy bias evolution, which we describe in Section \ref{sec:32} . 

In the clustering redshift technique, one relates the normalized redshift distribution of the unknown sample, $n_{\text{u}}(z')$,  to the angular cross-correlation, $w_{\text{ur}}$, between the unknown sample and a narrow redshift slice of the reference sample.  In the limit of linear scale-independent biasing, $n_{\text{u}}(z')$ and $w_{\text{ur}}$ are related by:

\begin{equation}
 w_{\text{ur}}(\theta) = \int dz'  n_{\text{u}}(z') n_{\text{r}}(z') b_{\text{u}}(z')b_{\text{r}}(z')w_{\text{mm}}(\theta,z')
\label{crosscorr}
\end{equation}

\noindent where $n_{\text{r}}$ is the normalized redshift distribution of the reference samples, $b_{\text{u}}$ and $b_{\text{r}}$ are the galaxy biases of the unknown and reference samples, and $w_{\text{mm}}$ is the two-point correlation function of the full matter distribution.

For our measurement of $ w_{\text{ur}}(\theta)$, we implement the method used in \cite{Schmidt2013}. The method counts unknown galaxies in annuli around each reference galaxy bounded by comoving scales $r=(1+z)D_{\text{A}} \theta$ from $r_{\text{min}}$ to $r_{\text{max}}$, where $D_{\text{A}}$ is the angular diameter distance to create a single-value estimate of the cross-correlation between two samples. For our analysis we set $r_{\text{min}}=500$ kpc and $r_{\text{max}}=1500$ kpc.\footnote{We note that the scales mentioned above include the non-linear regime of density fluctuations which puts into question the assumption of a linear scale-independent galaxy bias model in Equation \ref{crosscorr}. \cite{xcorrtechnique} studies a variety of scales in simulations for the fiducial method used here, and finds that implementing a linear bias model on these scales does not significantly impact the accuracy of the method. Our results in Section \ref{sec:sdss_tests}, where we test the method on redMaGiC galaxies with spectroscopic redshifts, also supports the conclusion that using these scales does not significantly impact the accuracy of the results.} This choice of scales is based on the work in \cite{xcorrtechnique} where the impact of scales on the method is analyzed in simulations. The counted galaxies are also inverse-weighted by distance which improves the S/N ratio of the measurement \citep{Schmidt2013}. For the single-value cross-correlation, we use the estimator from \cite{DavisPeebles1983}:

\begin{equation}
\label{DavisPeeblesestimator}
\bar{w}_{\text{ur}} =\frac{N_{\text{Rr}}}{N_{\text{Dr}}}\frac{\int_{r_{\text{min}}}^{r_{\text{max}}}dr' W(r')\left[D_{\text{u}}D_{\text{r}}(r')\right]}{\int_{r_{\text{min}}}^{r_{\text{max}}}dr' W(r')\left[D_{\text{u}}R_{\text{r}}(r')\right]}-1,
\end{equation}

\noindent where $\bar{w}_{\text{ur}}$ is the single-value cross-correlation, $({D_{\text{u}}D_{\text{r}}(r')})$ and $({D_{\text{u}}R_{\text{r}}(r')})$ are the numbers of data-data and data-random pairs in different angular bins set by $r'$, $N_{\text{Rr}}$ and $N_{Dr}$ are the numbers of galaxies in the reference sample and reference randoms and $W(r')$ is a weighting function set to $1/r'$ as part of the method in \cite{Schmidt2013}. Equation \ref{DavisPeeblesestimator} uses a random catalog for the reference sample as mentioned in Section \ref{sec:sdss}. Only the randoms for the reference sample are used to be consistent with the analysis of \cite{xcorrtechnique}. Here and below, all our cross- and auto-correlations (e.g., Equations \ref{autocorrelation}-\ref{autocorrelation2}) will use this single-value estimate defined in Equation \ref{DavisPeeblesestimator} for each pair of unknown and reference redshift-binned samples.

In practice, we will evaluate Equation \ref{DavisPeeblesestimator} in narrow discrete redshift bins of the reference sample ($dz=0.01$ or 0.02) centered at $z$. In each reference bin, the normalized $n_{\text{r}}(z)$ is just 1, though $n_{\text{u}}(z)$ is still unknown, as $n_{\text{u}}$ goes over the full redshift range of the unknown sample of interest (which in this work is binned by $dz=0.15$). We can then invert Equation \ref{crosscorr} to obtain the number of galaxies in the unknown sample in each of these reference sample redshift bins: 

\begin{equation}
\label{menard2}
n_{\text{u}}(z) \propto  \bar{w}_{\text{ur}}(z)  \frac{1}{{b}_{\text{u}}(z)} \frac{1}{{b}_{\text{r}}(z)}
  \frac{1}{\bar{w}_{\text{mm}}(z)}
\end{equation}

\noindent where the barred quantities indicate we are now using the single-value estimator of Equation \ref{DavisPeeblesestimator}. Equation \ref{menard2} also assumes that the galaxy biases and the matter-matter correlation function are not changing significantly across the reference bin centered at $z$. The reference redshift bins are narrow, only $dz=0.01$ for the SDSS redMaGiC analysis and $0.02$ for the DES analysis. We use larger bins in the DES analysis to have more galaxies per bin and reduce statistical errors.

Alternative methods to the above are also studied in \cite{xcorrtechnique} but we use the preferred method identified in that work. The similar method from \cite{Menard2013} produces comparable results. Alternative estimators were also tested, such as using only randoms for the unknown sample instead of reference, a test that could reveal issues in the spectroscopic dataset. The Landy-Szalay estimator \citep{LandySzalay1993} was also tested in our analysis, which uses randoms for both samples of galaxies. We note that using randoms for the unknown sample would be a more acceptable option in our work than it was in \cite{xcorrtechnique} and \cite{xcorr}, where an accurate mask and random catalog was difficult to make for the weak lensing source galaxies in DES Y1. Our tests found that changing the estimators produces far less variance than the systematic errors described below, namely the effects of correcting for galaxy bias, so we decided to match the estimator used in \cite{xcorrtechnique} and \cite{xcorr} for consistency.

\subsection{Correcting for Galaxy Bias}
\label{sec:32}

Equation \ref{menard2} gives us a solution for calculating $n_{\text{u}}(z)$ with a few unknowns. In the calibration of weak lensing source galaxies in \cite{xcorr}, corrections for the galaxy bias are not part of the fiducial procedure. In the simulations studied in \cite{xcorrtechnique}, the galaxy bias redshift evolution of the sources is found to be quite complex and difficult to model. The redMaGiC galaxy bias evolution in the simulations (also studied here in Section \ref{sec:autocorrs}) is far more smooth in comparison, but it is only a minor correction in \cite{xcorrtechnique}. Instead of correcting for either galaxy bias directly, \cite{xcorrtechnique} assesses how much ignoring the galaxy biases changes the results in simulations. This is used to estimate systematic errors to the method, and \cite{xcorrtechnique} finds that the galaxy bias evolution of the source galaxies is the dominant systematic effect in their measurement. 

In contrast, in this work we do attempt to correct for these galaxy biases. This change in approach is warranted by both the higher S/N of this measurement between redMaGiC and spectroscopic galaxies, and by the accurate measurements available to assess the galaxy bias evolution of each sample. The effects of not correcting for galaxy bias for some datasets are demonstrated in the tests of Section \ref{sec:sdss_tests}. 

We can correct for the galaxy bias of the reference sample with the auto-correlation of the reference sample at different redshift bins:

\begin{equation}
\label{autocorrelation}
\bar{w}_{\text{rr}}(z)=b_{\text{r}}(z)^2 \bar{w}_{\text{mm}}(z)
\end{equation}

\noindent where each $z$ refers to a different reference redshift bin. This measurement can be done for each redshift bin of the reference sample. In principle, a similar correction using the auto-correlation for the unknown sample could also be used. The auto-correlation of the unknown is:

\begin{equation}
\label{autocorrelation2}
\bar{w}_{\text{uu}}(z)=b_{\text{u}}(z)^2 \bar{w}_{\text{mm}}(z) .
\end{equation}

\noindent Then following from Equation \ref{menard2}, we could solve for the redshift distribution with:

\begin{equation}
\label{bothauto}
n_{\text{u}}(z) \propto \frac{\bar{w}_{\text{ur}}(z)}{\sqrt{\bar{w}_{\text{rr}}(z) \bar{w}_{\text{uu}}(z)}} .
\end{equation}

\noindent However, the estimates of $\bar{w}_{\text{uu}}$ in the narrow redshift bins ($dz=0.02$ for DES) are noisy, and tests using Equation \ref{bothauto} directly on the SDSS redMaGiC sample with spec-z show that this approach leads to biased results for $n_{\text{u}}(z)$. To reduce the impact of the noise, we assume $w_{\text{uu}}$ evolves monotonically with redshift and approximate it with a simple power law. The assumption of passive evolution of the galaxy bias is supported theoretically (e.g., \cite{Tegmark96bias}) as well as by the measured galaxy bias of redMaGiC at larger scales in \cite{wthetapaper}. We use a similar power law parameterization to \cite{DAVIS}:

\begin{equation}
\label{powerlaw}
\sqrt{\bar{w}_{\text{uu}}(z)} \propto (1+z)^\gamma .
\end{equation}

\noindent This leads to our full estimator for the redshift distribution, $n_{\text{u}}(z)$:

\begin{equation}
\label{biascorrection}
n_{\text{u}}(z) \propto \frac{\bar{w}_{\text{ur}}(z)}{\sqrt{\bar{w}_{\text{rr}}(z)}} \frac{1}{(1+z)^\gamma} .
\end{equation}

The uncertainty in estimating the value of $\gamma$ to be used to get the correct redshift distribution, $n_{\text{u}}(z)$, is the largest systematic error in our analysis. Using Equation \ref{biascorrection} is in practice slightly different than directly using Equation \ref{bothauto}. In Section \ref{sec:autocorrs}, we use the auto-correlations of different redMaGiC datasets (across different surveys and luminosity cuts) over a large range of redshifts to make overall estimates of $\gamma$ for redMaGiC (Equation \ref{powerlaw}). This reduces the noise compared to directly using Equation \ref{bothauto} for any single redMaGiC dataset. We save further comments on estimating $\gamma$ for Section \ref{sec:systematics}.

An important caveat is that deriving Equation \ref{biascorrection} from Equation \ref{menard2} assumes $\bar{w}_{\text{uu}}$ is measured on true redshifts. (Otherwise, $b_{\text{u}}$ is a different quantity in Equation \ref{menard2} than in Equation \ref{autocorrelation2}.)  In practice, all of our samples of redMaGiC are selected by photometric redshift. When measuring the auto-correlation of a photometric redshift selected sample, the amplitude of the auto-correlation will be affected by the fact that the width of the true galaxy distribution in redshift space will be wider than if the galaxies were binned by their true redshifts. This effect will lower the amplitude of the auto-correlation, and could change the inferred redshift evolution of the galaxy bias represented by $\gamma$ in Equation \ref{biascorrection}. As shown in \cite{xcorrtechnique} (Appendix B), this error due to photo-z can be corrected with:

\begin{equation}
\label{photozcorr}
\bar{w}_{\text{uu}}(z) \propto \bar{w}_{\text{uu},\text{pz}}(z)  \frac{\int{N_{\text{spec}}(z)^2 dz}}{\int{N_{\text{pz}}(z)^2 dz}} 
\end{equation}

\noindent where $N(z)$ is the spectroscopic galaxy distribution of the unknown sample with the subscripts indicating whether binned by spectroscopic redshift or photometric redshift measurements, $\bar{w}_{\text{uu}}(z)$ is the auto-correlation of redMaGiC absent of photo-z effects (i.e., what you would measure selecting objects by spectroscopic redshifts), and $\bar{w}_{\text{uu},\text{pz}}(z)$ is the auto-correlation you measure on photo-z selected bins. This correction thus still needs spectroscopic information to assess the true distribution of galaxies when binned by photo-z, $N_{\text{pz}}(z)$. Spectroscopic information is needed in principle for $N_{\text{spec}}(z)$ as well, though in practice it is usually flat across a small redshift bin. We achieve this by measuring $N_{\text{pz}}(z)$ on small subsamples of our redMaGiC galaxies that have spectroscopic redshifts. If this subsample is not representative of $N_{\text{pz}}(z)$ for the entire sample of redMaGiC, this could lead to additional systematic errors. This is another reason we ultimately smooth out estimates of $\bar{w}_{\text{uu}}$ by a power law (Equation \ref{powerlaw}). We go into more detail on our estimates of $\bar{w}_{\text{uu}}$ and thus $\gamma$ both directly on true redshifts if available, and using Equation \ref{photozcorr} when only photometric redshifts are available in Section \ref{sec:systematics}.

Our last step in the clustering estimate is making a cut on the tails of the redshift distribution. This is necessary as the clustering redshifts method (Equations \ref{crosscorr}-\ref{menard2}), can be noisy and potentially biased in the tails of the redshift distribution where the S/N is low. In the tests on redMaGiC with spectroscopic measurements in Section \ref{sec:sdss_tests}, it is clear that the recovery of the true distribution in the tails is significantly biased compared to the higher amplitude parts of the distribution. We discuss this more in Section \ref{sec:44}. We cut the redshift distribution at $\pm 2.5 \sigma_{\text{u}}$ from the mean of the clustering redshift distribution estimate with $\sigma_{\text{u}}$ being the standard deviation of that estimate. \cite{xcorrtechnique} and \cite{xcorr} make a similar cut at $\pm 2 \sigma_{\text{u}}$. We opt to use more of the data, but cutting at $2 \sigma_{\text{u}}$ has a very minor effect on our results changing $\Delta z$ by about 0.001, well below our errors.

\subsection{Estimating Photometric Redshift Bias}
\label{sec:33}

The clustering method as described above provides a general estimate for the redshift distribution, $n_{\text{u}}(z)$, of a galaxy sample (Equation \ref{biascorrection}). We now shift the focus of the method to a more narrow goal of calibrating a photometric redshift distribution, $n_{\text{pz}}(z)$. DES Y1 analyses show that the most important feature of $n_{\text{pz}}(z)$ for the cosmological analyses is the mean redshift of the distribution (\cite{keypaper}, \cite{photoz}, \cite{methodpaper}, \cite{shearcorr}). We thus focus on calibrating a single mean bias, $\Delta z=\bar{z}_{\text{u}}-\bar{z}_{\text{pz}}$, of the photometric redshift distribution, $n_{\text{pz}}(z)$. Future work may include a more extensive calibration of the photometric distribution beyond this single shift estimation.

After estimating a value of $\gamma$ (see Section \ref{sec:autocorrs}) to complete our measurements of Equations \ref{powerlaw} and \ref{biascorrection}, and choosing a redshift range that we will use the clustering results on, we estimate $\Delta z$, the photometric redshift bias. We fit for $\Delta z$ by shifting the photometric redshift distribution to match the mean redshift, $\bar{z}_{\text{u}}$ of the clustering estimate, $n_{\text{u}}(z)$ in Equation \ref{biascorrection}, over the redshift range used. Specifically this is finding the $\Delta z$ that satisfies:

\begin{equation}
\label{meanfitting}
\frac{\int_{z_{\text{min}}}^{z_{\text{max}}} z \ n_{\text{pz}}(z-\Delta z) \ dz}{\int_{z_{\text{min}}}^{z_{\text{max}}} n_{\text{pz}}(z-\Delta z) \ dz}  =   \frac{\int_{z_{\text{min}}}^{z_{\text{max}}} z \ n_{\text{u}}(z) \ dz}{\int_{z_{\text{min}}}^{z_{\text{max}}} n_{\text{u}}(z) \ dz}
\end{equation}

\noindent where $n_{\text{pz}}$ is the photometric redshift distribution, and $z_{\text{min}}$ and $z_{\text{max}}$ are set by the clustering estimate to be $\bar{z}_{\text{u}} \pm 2.5 \sigma_{\text{u}}$ as mentioned previously. Our methodology assumes that the clustering estimate of the mean redshift, $\bar{z}_{\text{u}}$ is a more accurate estimate of the true mean than the mean of photo-z distribution, $\bar{z}_{\text{pz}}$. This assumption is tested on the SDSS redMaGiC sample with spectroscopic redshifts in the next section.

\section{Estimating Systematics}
\label{sec:systematics}

\subsection{Testing with a Spectroscopic SDSS redMaGiC Subsample}
\label{sec:sdss_tests}

To validate the methodology of Section \ref{sec:methods}, we test on a subsample of SDSS DR8 redMaGiC galaxies that have spectroscopic redshift measurements as mentioned in Section \ref{sec:datasets}. The results from these tests show the validity of the cross-correlation method, while also illuminating some important systematic issues. We note that while these subsamples with spec-z are a small percentage of the larger SDSS datasets, the subsamples are about six times larger than the DES redMaGiC samples in Stripe 82 in Bin 1 ($z \in [0.15,0.3]$), and three times larger than those in Bin 2 ($z \in [0.3,0.45]$).

\begin{figure*}
\begin{center}
\includegraphics[width=1.0 \textwidth]{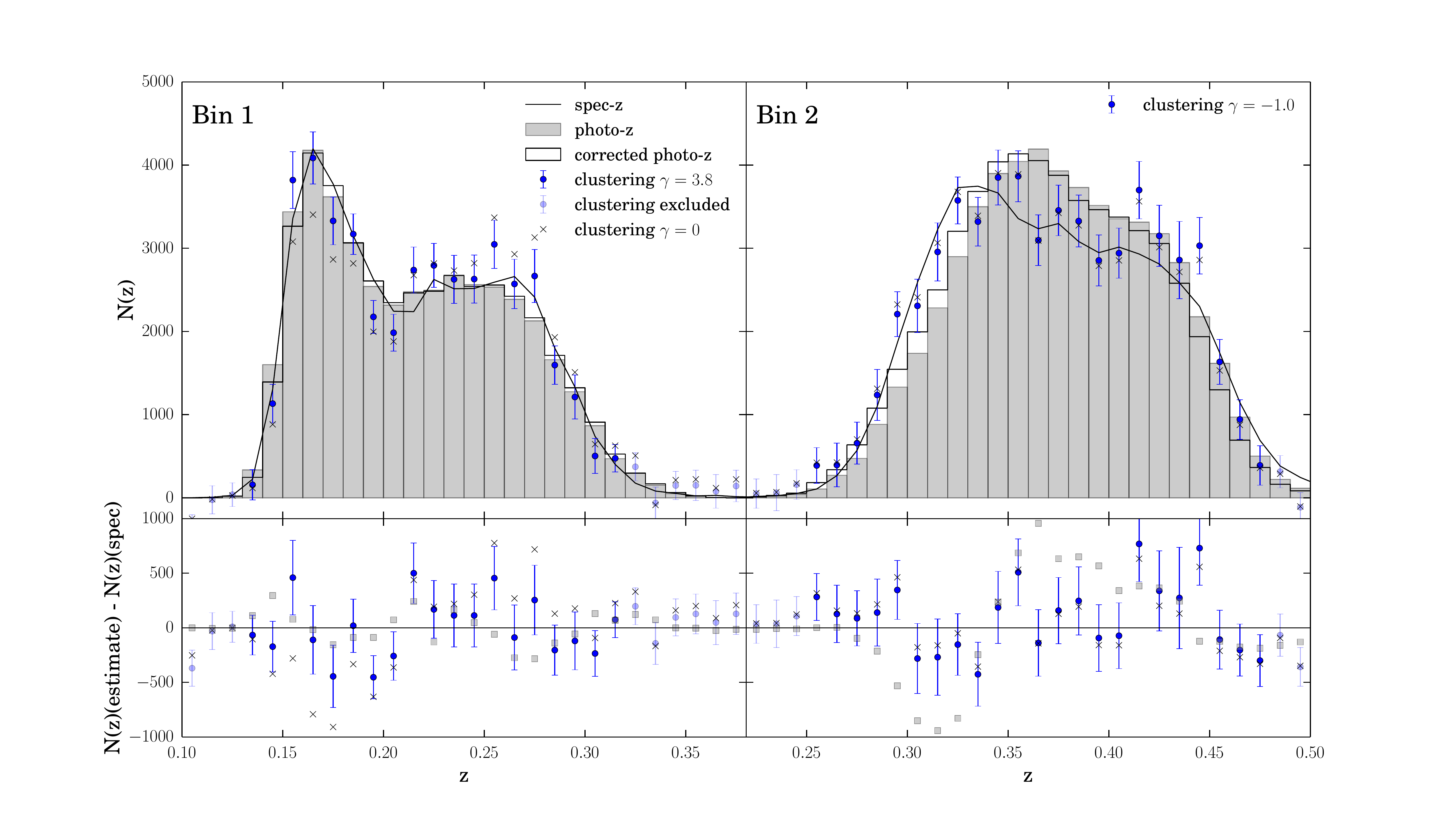}
\end{center}
\caption{(Top row) SDSS redMaGiC with spectroscopic redshifts, high-density sample, clustering redshift distributions ($n_{\text{u}}(z)$ in Equation \ref{biascorrection}), photo-z distribution ($n_{\text{pz}}(z)$ in Equation \ref{meanfitting}), shifted photo-z fit ($n_{\text{pz}}(z+\Delta z)$ in Equation \ref{meanfitting}) and true spec-z distribution. The blue points have the values of $\gamma$ that make the clustering estimate match the mean redshift of the spec-z (see Table \ref{table:sdss_spec_table}). The black x's are the clustering estimate with $\gamma=0$ (no bias correction).  The `clustering excluded' points are the clustering redshift estimates in the tails that are cut from analysis (see end of Section \ref{sec:32}). (Bottom row) The residuals (number of galaxies) comparing the clustering and photometric redshift distribution estimates with the spec-z distribution. The total residuals for the range where clustering is used are approximately for Bin 1: 5,400 for clustering, 2,700 for photo-z, Bin 2: 9,400 for clustering, 10,000 for photo-z.}
\label{fig:sdss_spectra4}
\end{figure*}

Importantly, we can use this sample to test the accuracy of the galaxy bias calibration method described in Section \ref{sec:32}. We measure the cross-correlation between the redMaGiC galaxies and the BOSS reference galaxies (Equation \ref{DavisPeeblesestimator}) and the auto-correlation of the reference galaxies and use Equations \ref{powerlaw}-\ref{biascorrection} to estimate the redshift distribution, starting with $\gamma=0$ (no bias correction). To assess the `true' bias correction for each of the two redshift bins of this sample, we fit $\gamma$ to be the value that makes the clustering-estimated mean redshift of the sample match the true mean redshift of the sample as measured by spectroscopic redshifts. These correctly bias-calibrated clustering results are shown in Figure \ref{fig:sdss_spectra4}. Also shown in that figure are the clustering distributions with no bias correction ($\gamma=0$). We discuss the large values for $\gamma$ in Figure \ref{fig:sdss_spectra4} more in Section \ref{sec:autocorrs} and Figure \ref{fig:lumplot}.

\begin{figure}
\begin{center}
\includegraphics[width=0.5 \textwidth]{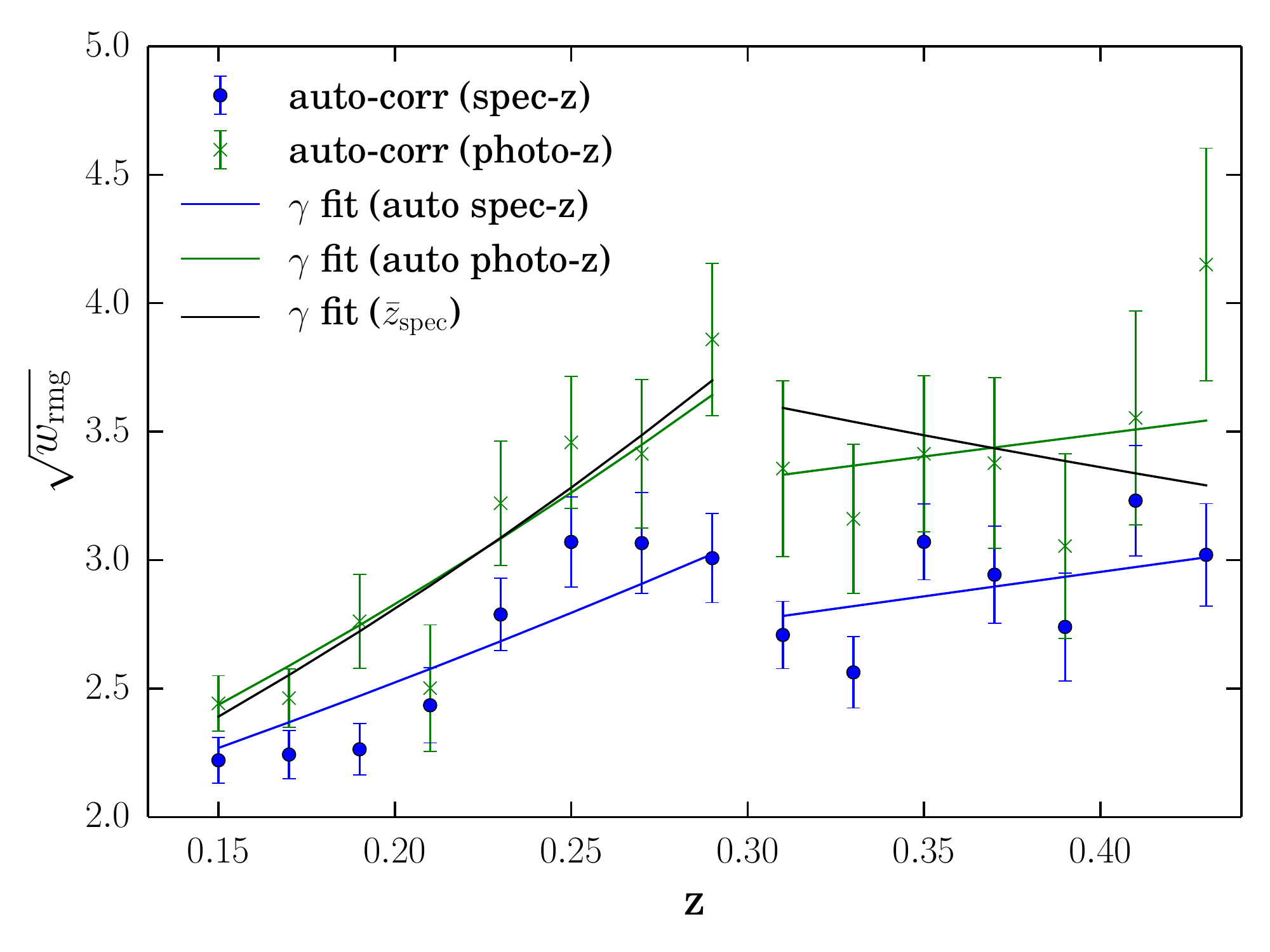}
\end{center}
\caption{Integrated auto-correlations (Equation \ref{DavisPeeblesestimator}) of SDSS redMaGiC with spectroscopic redshift samples. Shown are the auto-correlations of the same sample but on either the spectroscopic or photometric redshift measurements. The auto-correlations on the photo-z measurements of redMaGiC samples use the correction of Equation \ref{photozcorr}. We fit these auto-correlations in both Bin 1 ($z \in [0.15,0.3]$) and Bin 2 ($z \in [0.3,0.45]$) by $(1+z)^\gamma$ in Equation \ref{powerlaw}. The best fits are shown here, and their values are listed in Table \ref{table:sdss_spec_table}. We note that although the spec-z and photo-z auto-correlations are offset in amplitude, our methodology (i.e., Equation \ref{bothauto}) only depends on the redshift-dependence of the auto-correlations, parameterized by $\gamma$. The spec-z and photo-z auto-correlation fits are within $\delta \gamma=1$ of each other.}
\label{fig:autocorr_bias_sdss}
\end{figure}

We now test the accuracy of using auto-correlations on this sample to calibrate the bias correction by comparing the best-fit $\gamma$ values from the auto-correlations (Equation \ref{powerlaw}) to the value of $\gamma$ that yields the correct mean redshift. The auto-correlations are shown in Figure \ref{fig:autocorr_bias_sdss}, both as measured on spectroscopic redshifts and on photometric redshifts which requires also applying Equation \ref{photozcorr} to correct for photo-z effects on the auto-correlation. The resulting best fit $\gamma$'s to these auto-correlations, and the resulting photo-z biases from those $\gamma$'s are shown in Table \ref{table:sdss_spec_table}, along with the true photo-z biases and the $\gamma$'s that fit to those values.

\begin{table*}
\begin{center}
    \begin{adjustbox}{width=1.\textwidth}
    \begin{tabular}{|c|c|c|c|c|c|c|c|}
      \hline
      \multirow{3}{3cm}{}& \multicolumn{2}{c|}{Auto-Corr. on Spec-z} & \multicolumn{2}{c|}{Auto-Corr. on Photo-z} & \multicolumn{2}{c|}{Fit to $\bar{z}_{\text{spec}}$ (w/cut tails)} & Fit to $\bar{z}_{\text{spec}}$ (full bin) \\
      \hline
      Galaxy Sample & $\gamma$ & $\Delta z$ & $\gamma$ & $\Delta z$  & $\gamma$ & $\Delta z$ & $\Delta z$ \\
      \hline
      Bin 1 ($z \in [0.15,0.3]$) & $2.5 \pm 0.4$ & 0.0042 & $3.5 \pm 0.5$ & 0.0022 & $3.8 \pm 0.7$ & 0.0012 & 0.0010\\
      \hline
      Bin 2 ($z \in [0.3,0.45]$) & $0.9 \pm 0.6$ & -0.0079 & $0.7 \pm 0.9$ & -0.0076 & $-1.0 \pm 0.7$ & -0.0046 & -0.0026 \\      
      \hline
    \end{tabular}
  \end{adjustbox}
  \caption{Results of tests on the SDSS high-density redMaGiC sample with spec-z measurements. The top row indicates different methods of measuring the galaxy bias correction factor, $\gamma$, in Equation \ref{powerlaw}, and the resulting measured photo-z bias, $\Delta z$. The first two methods use auto-correlations of the sample on spec-z measurements (unavailable for our fiducial datasets), and on photo-z measurements while also using Equation \ref{photozcorr} to correct for photo-z effects. The last two columns (`fit to $\bar{z}_{\text{spec}}$') show what the true $\Delta z$ is, both over the redshift range used for clustering (`cut tails') and over the `full bin'. For the `cut tails', we show the value of $\gamma$ that makes the clustering estimate fit the correct $\bar{z}_{\text{spec}}$ over this redshift range.}
  \label{table:sdss_spec_table}
  \end{center}
\end{table*}

We can infer from these comparisons the relative accuracy of the auto-correlation method for correcting the galaxy bias evolution effects. Each of the four estimates of $\gamma$ (including the spec-z and photo-z auto-correlations) are within $\pm 2$ of the `true' $\gamma$ that fits the spec-z mean, or approximately within $\pm 0.004$ from the true photo-z bias, $\Delta z$. We also see that, although in Figure \ref{fig:autocorr_bias_sdss} the auto-correlations on spec-z and photo-z measurements of redMaGiC do not perfectly match up, they yield similar values of $\gamma$ compared to the overall scatter of $\gamma$ measurements on this subsample from the `true' values. The photo-z results actually match the true $\gamma$'s slightly better, but this is well within the errors of fitting $\gamma$ to the auto-correlations. In any case, the difference between photo-z and spec-z auto-correlations appears sub-dominant to the overall uncertainty of using auto-correlations to get the correct $\gamma$ and thus the correct mean redshift. Figure \ref{fig:autocorr_bias} (left panel) in the next section also supports this observation.

Another observation to note in Table \ref{table:sdss_spec_table} is  that when we cut the tails of the redshift distribution (see Section \ref{sec:33}) for the clustering estimate, we also change the true photo-z bias of the redshift range we are measuring in. This effect is seen in Table \ref{table:sdss_spec_table} by comparing $\Delta z$ fit to $\bar{z}_{\text{spec}}$ over the full bin to the same fit over the `cut tails' redshift range. The difference in $\Delta z$ is around $0.002$. Cosmological analyses usually depend on the photo-z bias over the full bin, but our measurements trace the photo-z bias over only the cut range. Since our measurements are primarily meant to aid cosmological analyses, this is another systematic error of our measurements.

In comparing the accuracy of clustering redshifts to photometric redshifts on this sample, we can see that e.g., the clustering method bias, |$\Delta z$ (pz auto) - $\Delta z_{\text{spec}}$ (cut tails)|, is more than a factor of two less than the photo-z bias, |$\Delta z_{\text{spec}}$ (cut tails)|, in each bin, as seen in Table \ref{table:sdss_spec_table}. On the other hand, in Bin 1, the residuals are about a factor of 2 less for photo-z than clustering, and are about equal for Bin 2 (Figure \ref{fig:sdss_spectra4}). Based on these results, and the uncertainty of having just two bins to test on, we can say little more than that the methods have roughly comparable accuracy for this sample. However, the full photometric samples of redMaGiC will primarily be fainter galaxies compared to this test sample with spec-z measurements, likely making their photo-z measurements less accurate. We confirm this in Section \ref{sec:photozerrors}. Also, the full samples will have more galaxies and thus smaller errors for the clustering method. The similar errors between photo-z and clustering on this bright sample of redMaGiC thus indicate that clustering is likely to be more accurate for the full samples.

\subsection{Galaxy Bias Evolution of redMaGiC}
\label{sec:autocorrs}

In this section we explore how we will calibrate the galaxy bias evolution systematic for the main redMaGiC samples. We could simply take the auto-correlations of e.g., the DES galaxies in Stripe 82, and apply the best-fit $\gamma$ with some uncertainty. However, the errors may be larger than when we tested the SDSS redMaGiC with spec-z estimates sample, since there are fewer galaxies. We fortunately have the ability to look at larger datasets to estimate the galaxy bias evolution of redMaGiC as well.

\begin{figure*}
\begin{center}
\includegraphics[width=1.0 \textwidth]{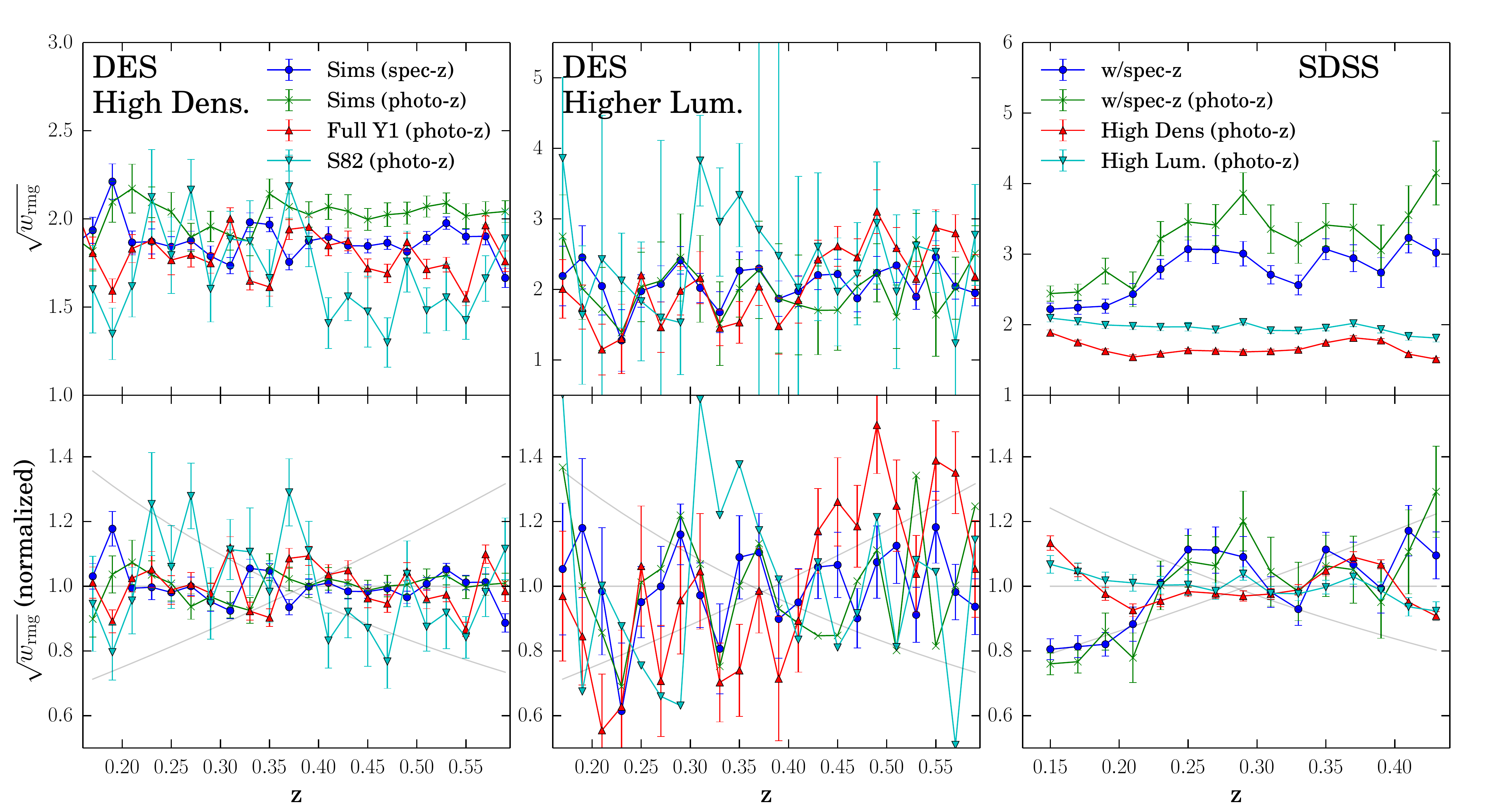}
\end{center}
\caption{(Top row) Integrated auto-correlations (Equation \ref{DavisPeeblesestimator}) of DES and SDSS redMaGiC samples. The photo-z redMaGiC samples use the correction of Equation \ref{photozcorr}. (Bottom row) The same auto-correlations normalized for better comparison between samples. Only the change in redshift matters as a systematic to the cross-correlation method, not the amplitude of the galaxy auto-correlation (Equation \ref{bothauto}). We fit the normalized auto-correlations by $(1+z)^\gamma$ in Equation \ref{powerlaw}. Shown in gray are lines for $(1+z)^\gamma$ with $\gamma=-2,0,2$ as examples, though we do fit across all possible values of $\gamma$. Each of the DES high-density samples and the SDSS high-density and high-luminosity samples are very consistent with $\gamma=0$. The DES higher-luminosity samples are much noisier due to fewer objects, so we also assume in our fiducial analysis $\gamma=0 \pm 2$. For the DES higher-luminosity sample, the full DES dataset is closer to $\gamma=2$, though the simulations and Stripe 82 data agree with $\gamma=0$. In the normalized DES higher-luminosity plot, the error bars for the simulations (photo-z) and the Stripe 82 data have been removed for clarity since they span the entire y-axis range of the plot. Also shown in the right panels again are the SDSS subsamples with spec-z measurements used in Section \ref{sec:sdss_tests}. Their significantly different bias evolution with redshift is apparent compared to the other redMaGiC samples (see Figure \ref{fig:lumplot} for more discussion on this).}
\label{fig:autocorr_bias}
\end{figure*}

In Figure \ref{fig:autocorr_bias}, we show the auto-correlations for a number of redMaGiC samples from DES and SDSS. Included are measurements of redMaGiC in the Buzzard simulations \citep{DeRose2017} as well as the full redMaGiC sample used in the DES Year 1 analyses in \cite{wthetapaper} and \cite{keypaper}. These auto-correlations are plotted along with the auto-correlations of DES redMaGiC in Stripe 82, the only sample we use for the cross-correlations. For SDSS, we plot the auto-correlations of the main samples (high-density and high-luminosity) as well as for the subsamples with spec-z again as seen in Figure \ref{fig:autocorr_bias_sdss}.

\begin{figure}
\begin{center}
\includegraphics[width=0.5 \textwidth]{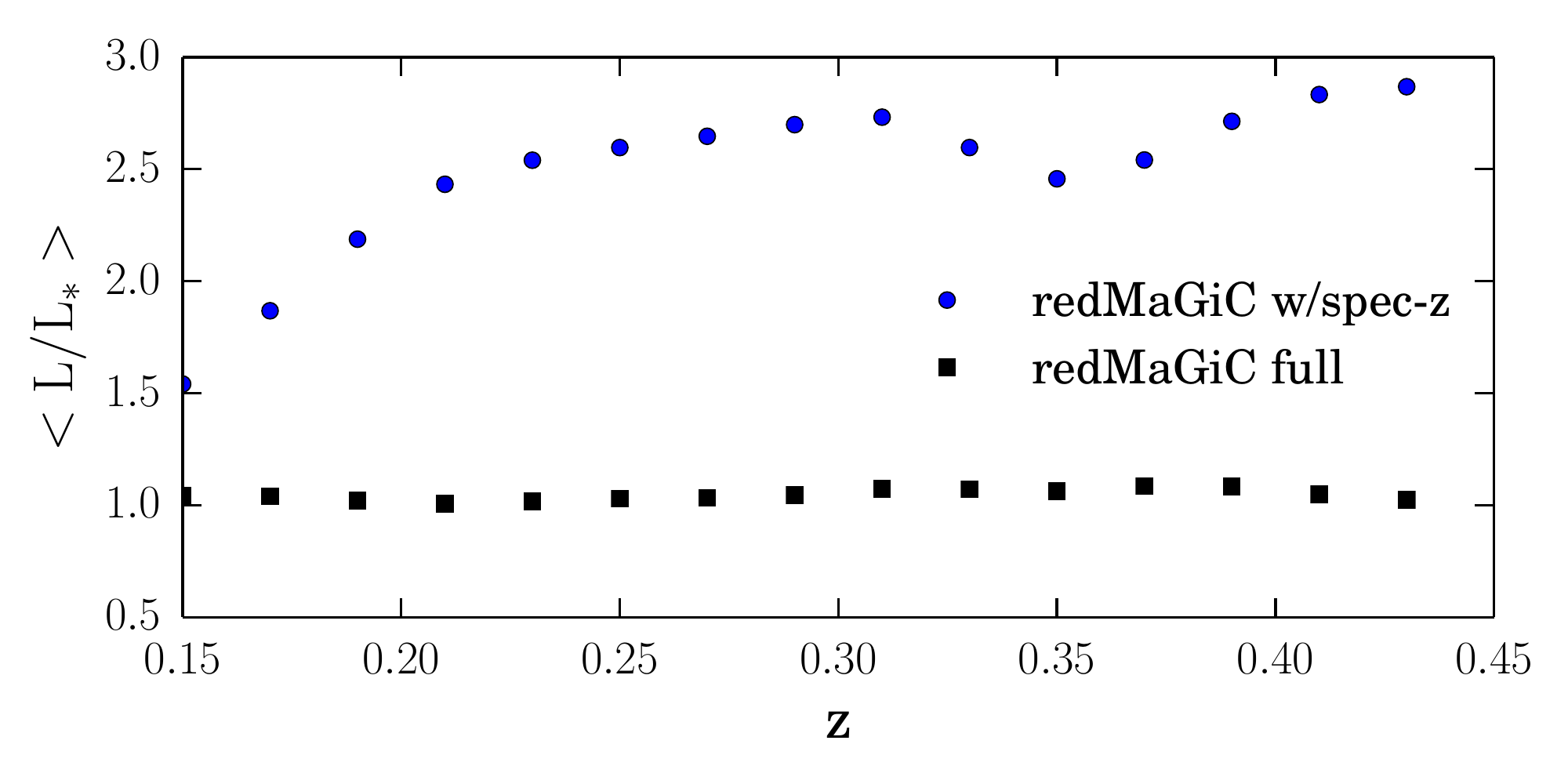}
\end{center}
\caption{The mean luminosity as a function of redshift for the SDSS DR8 redMaGiC samples. Shown is the full high-density sample, and the subsample that has spectroscopic redshifts. The reference luminosity, $L_*$ is described in Section \ref{sec:des}. Since galaxies with large apparent brightness are preferentially selected for spectroscopic redshifts, the mean luminosity of the spectroscopic sample increases with redshift. More luminous galaxies tend to have larger galaxy bias values. This likely explains the significant difference in galaxy bias evolution with redshift of these samples in Figure \ref{fig:autocorr_bias}, right panel.}
\label{fig:lumplot}
\end{figure}

Notably, we see in Figure \ref{fig:autocorr_bias}, that the SDSS sample with spec-z measurements used in Section \ref{sec:sdss_tests} has distinctly larger galaxy bias evolution with redshift than any of the other samples. This is likely due to luminosity bias, as more luminous galaxies are known to have stronger clustering and larger galaxy bias values (e.g., \cite{zehavi}, \cite{coupon2012}, \cite{Crocce16}). As seen in Figure \ref{fig:lumplot}, the sample with spec-z measurements increases in luminosity with redshift much more strongly than the full SDSS redMaGiC sample. This is a selection effect as galaxies targeted for spectroscopic measurements typically have larger apparent brightness. High redshift galaxies with spectra will thus be preferentially intrinsically brighter. 


In contrast to the SDSS sample with spec-z, the auto-correlations of the main redMaGiC samples in Figure \ref{fig:autocorr_bias}, and thus their galaxy biases, shows little evolution as a function of redshift. Each of the DES high-density samples, and both of the SDSS full photometric samples fit well with $\gamma=0$, a flat line. For the DES higher-luminosity samples, which are considerably noisier, the simulations and the Stripe 82 data are consistent with $\gamma=0$, though the full footprint DES sample shows a larger bias evolution, with a best fit of $\gamma \approx 2$.

Using the results of Figure \ref{fig:autocorr_bias}, and the tests on the SDSS sample with spec-z measurements (Figure \ref{fig:sdss_spectra4} and Table \ref{table:sdss_spec_table}), we decide to model the galaxy bias evolution of our main samples in DES and SDSS as a power law of $(1+z)^\gamma$ with $\gamma=0 \pm 2$. In practice this means we set $\gamma=0$ in Equation \ref{biascorrection} for our fiducial results and measure the difference in our estimated photo-z bias when setting $\gamma=2$ and $\gamma=-2$. Given the size of our redshift bins ($dz=0.15$), the $\pm 2$ in $\gamma$ always yields a scatter of approximately $\pm 0.004$ in $\Delta z$. 

The choice of $\delta \gamma=2$ is generally larger than the statistical uncertainty of determining $\gamma$ in either of the above tests. However, the choice of a broad prior of $\gamma=0 \pm 2$ reflects an uncertainty in how well the tests capture the errors of the bias calibration on the data. The SDSS redMaGiC subsample with spec-z's (Section \ref{sec:sdss_tests}) is a large sample we can test our full methodology on, but that redMaGiC sample has a significantly different bias evolution than the full photometric samples. We can estimate galaxy bias evolution on real photometric redMaGiC samples, or the simulated ones with auto-correlations, but we do not directly test how well those auto-correlations correctly calibrate the cross-correlations. Future work with simulated redMaGiC and simulated spectroscopic surveys may yield a more precise estimate of the uncertainties in our galaxy bias calibration method on samples more similar to our data than the SDSS redMaGiC subsample with spec-z's.

We also note that by $\chi^2/\text{dof}$, the $(1+z)^\gamma$ models are not always good fits of the auto-correlations of Figure \ref{fig:autocorr_bias}. The choice of using the power law (Equation \ref{powerlaw}) along with the broad prior of $\delta \gamma=2$ is still appropriate given the uncertainties in the method not captured in the statistical error bars of the figure, such as the uncertainty in the photo-z correction (Equation \ref{photozcorr}) which we do not directly estimate other than the comparison of methods in Table \ref{table:sdss_spec_table}. Noisy points from using this correction were why using the power law formalism was more accurate in testing with the SDSS subsample with spec-z than using auto-correlations directly. 

\subsection{Galaxy Bias Evolution of the Reference Spectroscopic Galaxies}
\label{sec:43}

The galaxy bias evolution of the reference sample from SDSS DR12 will also impact the cross-correlation of the reference and unknown samples. This effect is accounted for by the auto-correlation of the reference sample in Equation \ref{biascorrection}. 

\begin{figure}
\begin{center}
\includegraphics[width=0.5 \textwidth]{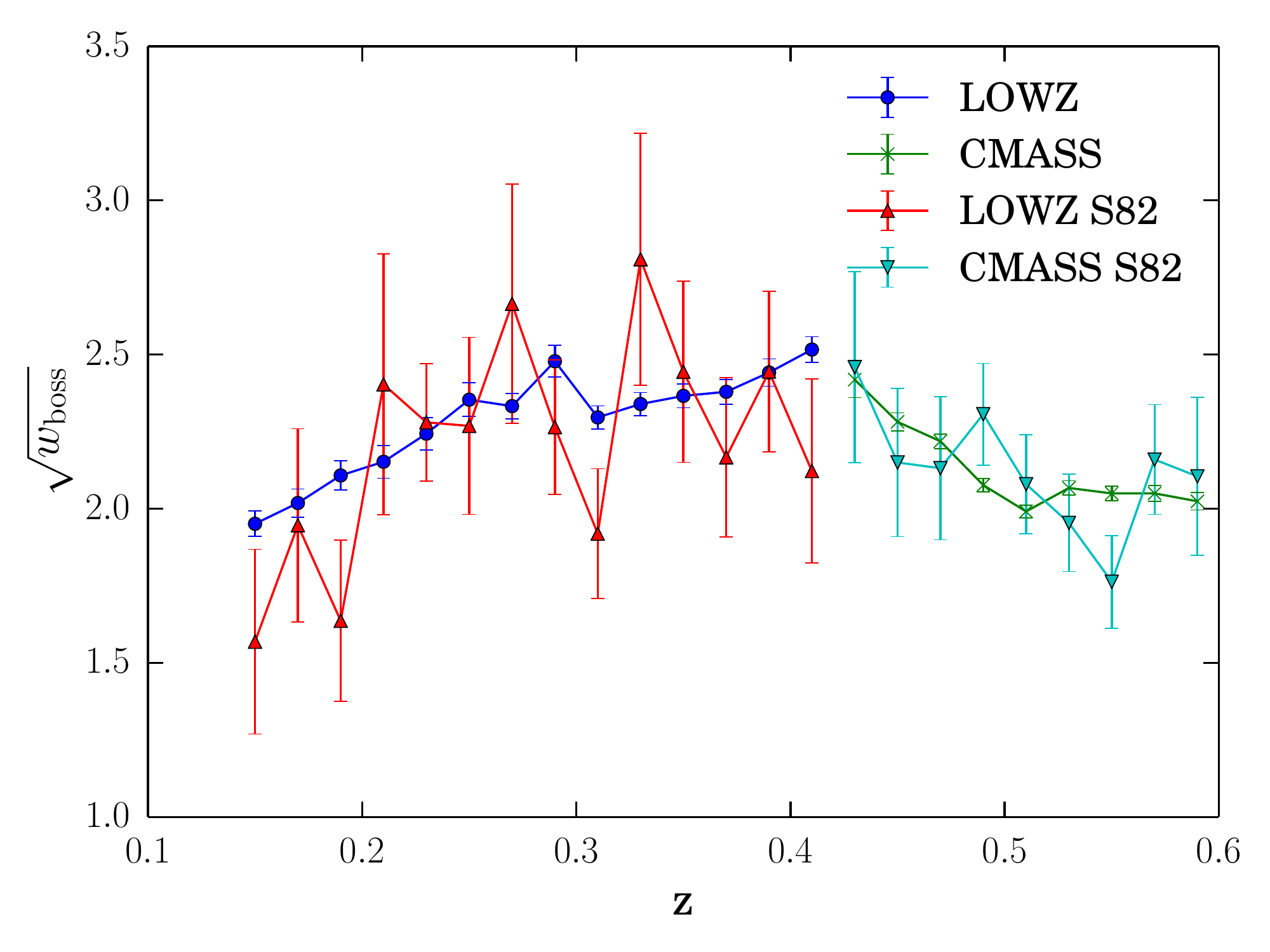}
\end{center}
\caption{The square root of the integrated auto-correlation (Equation \ref{DavisPeeblesestimator}) of each of the reference samples from SDSS DR12, which is broken up into the LOWZ and CMASS surveys. The auto-correlations of the full sample are statistically consistent with the auto-correlations on the reference sample just within Stripe 82, but with less noise ($\chi^2/\text{dof} \leq 1$ for both LOWZ and CMASS).}
\label{fig:refautocorr_bias}
\end{figure}

Similar to the previous discussion on how to treat the redMaGiC galaxy bias evolution, we again have an option in some cases to look at more data than just the samples directly used in the cross-correlations. For the measurements on DES redMaGiC, we can only use reference galaxies in about 124 $\text{deg}^2$ of Stripe 82, a far smaller sample than the total BOSS dataset used on the cross-correlations with SDSS redMaGiC. Figure \ref{fig:refautocorr_bias} shows the similarity between the auto-correlations of the reference sample over the entire SDSS footprint and the sample that overlaps DES in Stripe 82. They are statistically consistent, with $\chi^2/\text{dof} \leq 1$ for both LOWZ and CMASS. We note that similar estimates of the bias evolution of the samples in Figure \ref{fig:refautocorr_bias} have been made (\cite{reid2014}, \cite{manera2015}, \cite{salazar2017}). In the calibration of the redshift distributions for DES redMaGiC, we use the auto-correlation of the reference sample over the full BOSS dataset in Equation \ref{biascorrection}, rather than only the Stripe 82 galaxies in order to minimize noise. The difference in $\Delta z$ based on which reference sample we use for the auto-correlations is minimal, only 0.000-0.002 in $\Delta z$ for different datasets. Given the consistency between the full and S82 BOSS datasets, we choose not to include this difference as a systematic error, as it is likely due to statistical noise. 

\subsection{Influence of Tails of Redshift Distribution}
\label{sec:44}

The last systematic we attempt to account for are errors associated with cutting the tails of the redshift distribution. As discussed in \cite{xcorrtechnique}, and also seen in Figure \ref{fig:sdss_spectra4}, the cross-correlation method can be biased and noisy in the tails, where there is little signal. The bias may be due to magnification, as discussed in \cite{xcorrtechnique}. In Bin 2 in Table \ref{table:sdss_spec_table}, the reduced redshift range has a true photo-z bias that differs by 0.002 from the full bin's true photo-z bias. For each sample, we assign a minimum systematic error related to the tails of 0.002 based on those tests. We test further the impact of cutting the tails by checking how much the photo-z bias, $\Delta z$, changes when we change the redshift range used between 2 and 2.5 $\sigma_{\text{u}}$ (see Section \ref{sec:33}). In most cases, changing the redshift range in this way causes the $\Delta z$ to change by 0.001-0.003. In the few samples that have a deviation of about 0.003, we use this larger value for the systematic error related to the tails instead of 0.002. We include this error in our systematic errors reported in Tables \ref{table:des_final_table}-\ref{table:sdss_final_table}. 

\section{Results}
\label{sec:results}

We follow our procedure of Section \ref{sec:methods} to measure the cross-correlations of BOSS/SDSS DR12 spectroscopic galaxies and redMaGiC galaxies to get estimates of the redMaGiC redshift distribution and photometric redshift bias. We use the choice of no redshift evolution of the quantity $b_{\text{rmg}} \sqrt{\bar{w}_{\text{mm}}}$, setting $\gamma=0 \pm 2$ in Equation \ref{powerlaw} as detailed in Section \ref{sec:systematics}. Changing $\gamma$ from 0 to 2 or -2 always shifts $\Delta z$ by approximately $\pm 0.004$, so we assign 0.004 as the uncertainty on $\Delta z$ due to $\gamma$. We also assign an error of 0.002-0.003 depending on the sample based on the discussion of effects caused by cutting the tails of the redshift distribution from being used in the clustering estimate in Section \ref{sec:44}. Adding these errors in quadrature, we get for all of our main samples, a systematic error of $\pm 0.005$ in $\Delta z$. We add to this in quadrature the statistical error of the cross-correlation and auto-correlation of the reference sample (Equation \ref{biascorrection}) as measured by 100 jackknife samplings across the area of each sample. These statistical errors in $\Delta z$ range from 0.001 in SDSS to 0.010 in the DES higher-luminosity samples. The systematic, statistical and total uncertainties for each redMaGiC sample are shown in Tables \ref{table:des_final_table}-\ref{table:sdss_final_table}.

\subsection{DES redMaGiC Results}
Our main sample of interest is the Dark Energy Survey Year 1 redMaGiC catalogs, as they play a role in several other analyses, especially the high-density sample. As mentioned in Section \ref{sec:datasets}, our measurement is limited by the fact that the sample of DES redMaGiC galaxies in the overlapping region with BOSS is only about $10\%$ the size of the main redMaGiC samples used in the other Year 1 analyses. This significantly increases the statistical uncertainty of our measurements compared to those on SDSS redMaGiC. To improve S/N, we bin the reference sample in these measurements by 0.02 in redshift rather than by 0.01 which we do in the SDSS redMaGiC measurements. As mentioned in Section \ref{sec:systematics}, we use the full sample of BOSS spectroscopic galaxies for the auto-correlation of the reference sample in Equation \ref{biascorrection}. 

\begin{figure*}
\begin{center}
\includegraphics[width=1.0 \textwidth]{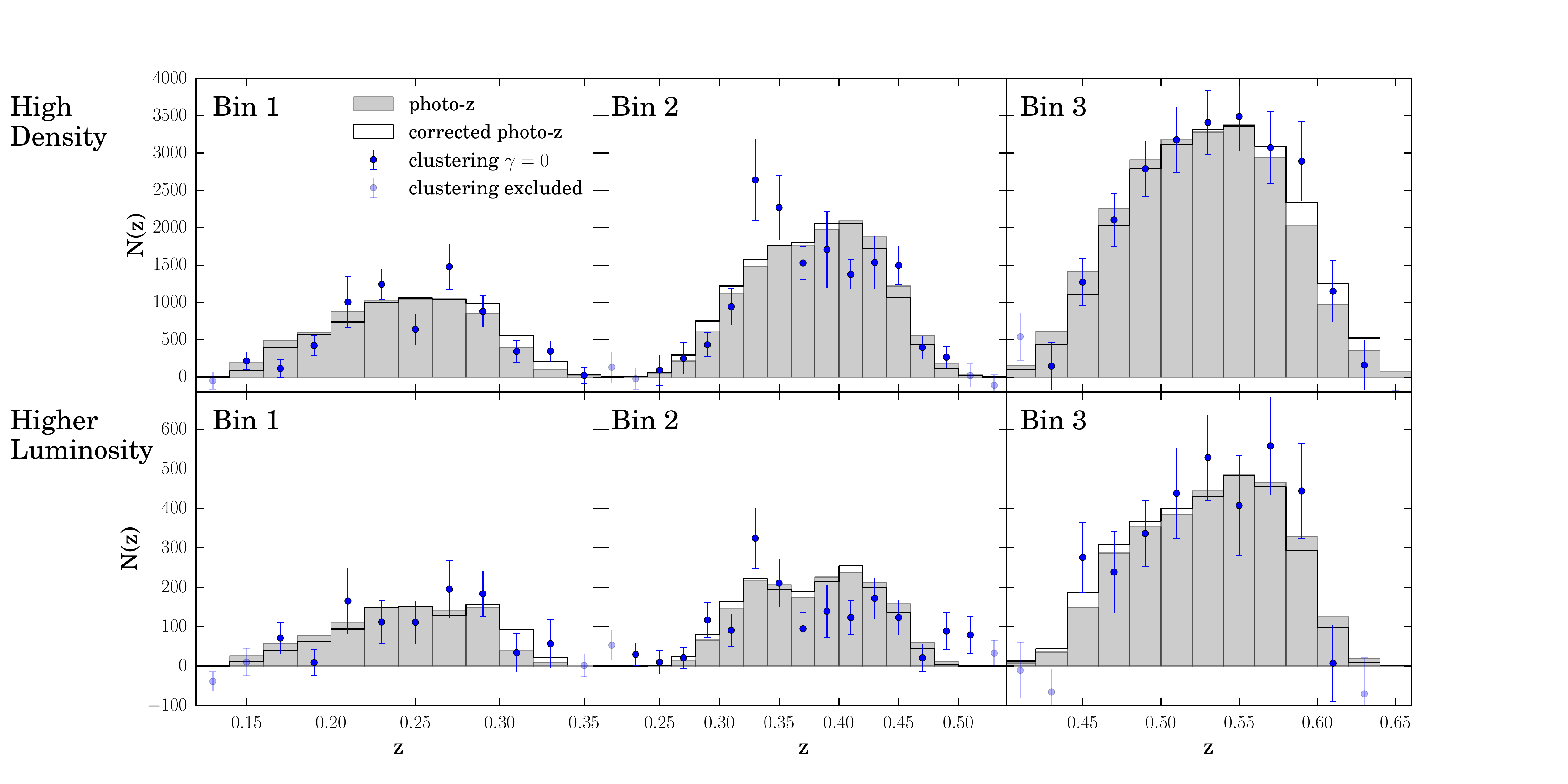}
\end{center}
\caption{DES Year 1 redMaGiC clustering-estimated redshift distributions and shifted photo-z fits.}
\label{fig:des6}
\end{figure*}

Our results\footnote{We note that v1 of the DES cosmology paper \cite{keypaper} used slightly different values than those shown here due to a late change in procedure when the cosmological analysis was close to completion. The cosmological impact between the previous measured biases and the current ones was negligible. The updated \cite{keypaper} now uses the current values shown in this work.} for both DES samples are shown in Table \ref{table:des_final_table} and Figure \ref{fig:des6}. The results overall show relatively small photometric redshift biases in DES redMaGiC, typically within 1 $\sigma$ of zero bias. Of note, the biases shown did not impact the cosmological results of \cite{keypaper} compared to having zero photo-z bias with similar uncertainty. The biases do seem somewhat larger than the estimation in \cite{redmagicSV} of a median bias of 0.005 though. Overall, there seems to be broad consistency between the estimates of photo-z bias in each redshift bin across the different science samples, DES high-density and higher-luminosity, and SDSS high-density and high-luminosity (Section \ref{sec:sdss}). The DES high-density and DES higher-luminosity are somewhat different in Bin 3 ($z \in [0.45,0.6]$) though.

We note the possibility of stronger galaxy bias evolution in the DES higher-luminosity samples. Our fiducial results in Table \ref{table:des_final_table} are computed with $\gamma=0 \pm 2$. In the auto-correlations of Figure \ref{fig:autocorr_bias}, the full footprint DES higher-luminosity sample is not fit well with this model. Its auto-correlation fit is $\gamma=2 \pm 0.5$. However, the DES higher-luminosity samples in the Stripe 82 data and the simulations are both fit well by $\gamma=0$, with fits of $\gamma=-0.1 \pm 0.9$ and $\gamma=0.0 \pm 0.3$. If $\gamma=2$ is a more correct calibration of this sample, it would move our fiducial photo-z biases, $\Delta z$, by -0.004. It is unclear if this discrepancy is due to noise, an issue in the simulations, or an issue in the methodology such as the application of the photo-z correction for the auto-correlation in Equation \ref{photozcorr}. It is plausible that this brighter sample would have more significant galaxy bias evolution, similar to the SDSS sample with spec-z. The DES higher-luminosity sample also has the largest statistical errors of any of the samples we study. Future work with more data may get a better estimate of this sample and how to calibrate its bias evolution. 

We note the smaller uncertainties in our analysis compared to the other cross-correlation method papers used on DES Y1 data, \cite{xcorr} and \cite{xcorrtechnique}, which used redMaGiC as a reference sample to calibrate the weak lensing source galaxies used in DES analyses. There are a few factors that clearly contribute to this better precision. Both the reference sample and the unknown sample have more accurate redshifts (comparing BOSS spectroscopic galaxies to redMaGiC, and redMaGiC to the weak lensing source galaxies). The bias evolution of the weak lensing source galaxies is larger and more complex than redMaGiC (\cite{xcorrtechnique}). Finally, the redshift bins for redMaGiC are smaller in redshift range than the weak lensing source galaxy bins, reducing impact of bias evolution across a bin.

\begin{table}
\begin{center}
    \begin{adjustbox}{width=0.5 \textwidth}
    \begin{tabular}{|l|c|c|c|}
      \hline
      DES redMaGiC Sample & $\Delta z$ & $\delta \Delta z$(syst) & $\delta \Delta z$(stat)\\
      \hline
      High-density Bin 1 ($z \in [0.15,0.3]$) & $0.008 \pm 0.007$ & 0.005 & 0.005 \\
      \hline
      High-density Bin 2 ($z \in [0.3,0.45]$) & $-0.005 \pm 0.007$ & 0.005 & 0.005 \\      
      \hline
      High-density Bin 3 ($z \in [0.45,0.6]$) & $0.006 \pm 0.006$ & 0.005 & 0.004 \\      
      \hline
      Higher-lum.  Bin 1 ($z \in [0.15,0.3]$) & $0.010 \pm 0.011$ & 0.005 & 0.010 \\
      \hline
      Higher-lum. Bin 2 ($z \in [0.3,0.45]$) & $-0.004 \pm 0.010$ & 0.005 & 0.008 \\      
      \hline
      Higher-lum. Bin 3 ($z \in [0.45,0.6]$) & $-0.004 \pm 0.008$ & 0.005 & 0.006 \\      
      \hline
    \end{tabular}
  \end{adjustbox}
  \caption{Main results for the redMaGiC photometric redshift biases, $\Delta z$, in DES Year 1 data.}
  \label{table:des_final_table}
  \end{center}
\end{table}

\subsection{SDSS redMaGiC Results}
\label{sec:sdss}
\begin{figure*}
\begin{center}
\includegraphics[width=1.0 \textwidth]{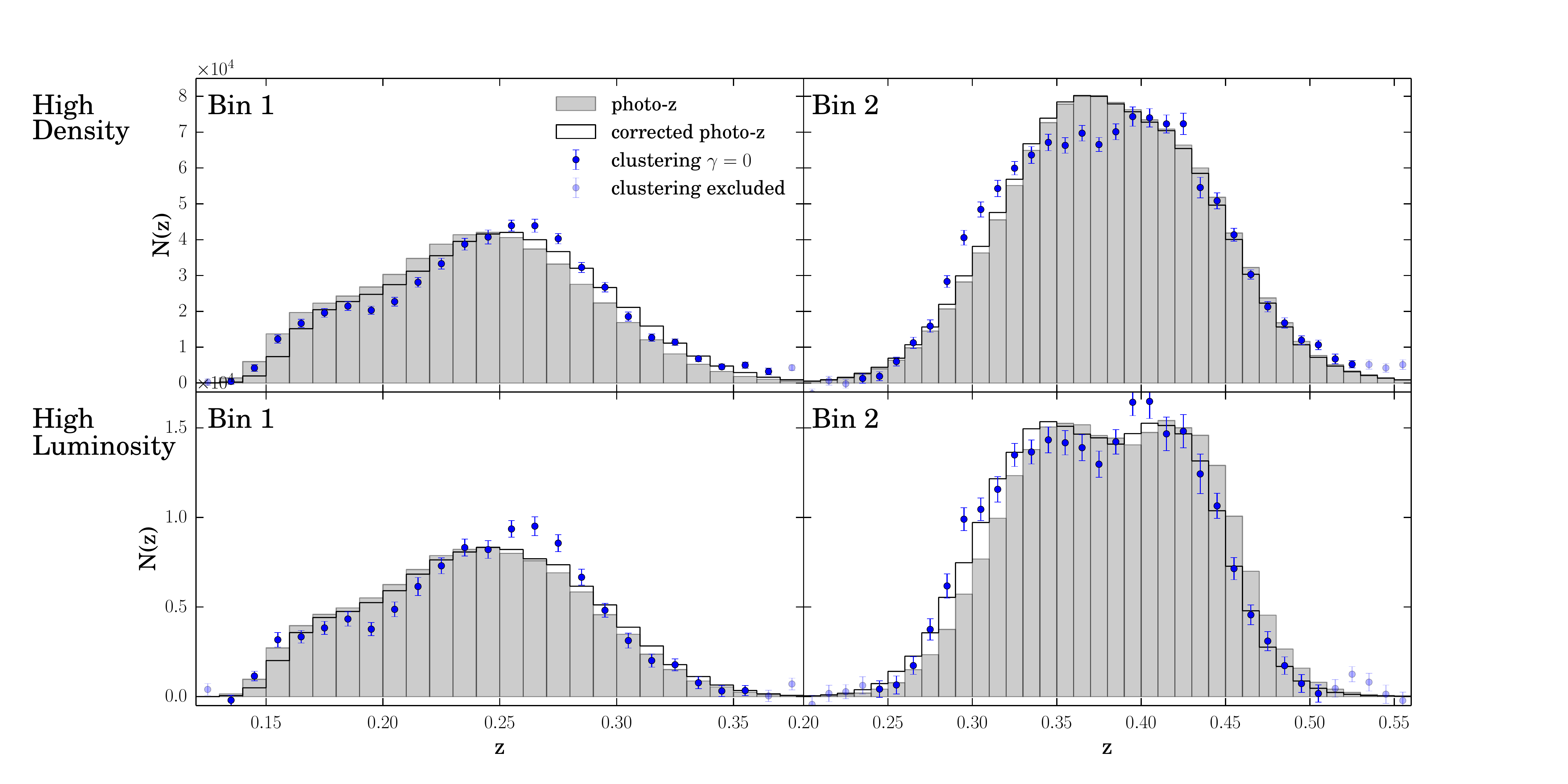}
\end{center}
\caption{SDSS redMaGiC clustering-estimated redshift distributions and shifted photo-z fits.}
\label{fig:sdss4}
\end{figure*}

While our main sample of scientific interest is the DES redMaGiC, studying the SDSS redMaGiC allows us to study a far larger sample with more constraining power. The full SDSS redMaGiC samples are about 15 times larger than the subsample with spec-z, and about 50 times larger than the DES sample in Stripe 82 that we can use for cross-correlations (comparing high-density samples). To compare with the DES results, we use the same binning as DES Bin 1 ($z \in [0.15,0.3]$) and Bin 2 ($z \in [0.3,0.45]$). The SDSS redMaGiC catalogs do not cover the DES Bin 3 ($z \in [0.45,0.6]$). These larger catalogs have statistical errors on $\Delta z$ of only around 0.001 for both the high-density and high-luminosity samples on these bins in SDSS. Our results for the high-density and high-luminosity SDSS samples are shown in Table \ref{table:sdss_final_table} and Figure \ref{fig:sdss4}. Overall, the photo-z biases appear similar to the DES photo-z biases across the $dz=0.15$ bins of our analysis, though we explore this in more detail in Section \ref{sec:photozerrors}.

\begin{table}
\begin{center}
    \begin{adjustbox}{width=0.5 \textwidth}
    \begin{tabular}{|l|c|c|c|}
      \hline
      SDSS redMaGiC Sample & $\Delta z$ & $\delta \Delta z$(syst) & $\delta \Delta z$(stat)\\
      \hline
      High-density Bin 1 ($z \in [0.15,0.3]$) & $0.008 \pm 0.005$ & 0.005 & 0.001 \\
      \hline
      High-density Bin 2 ($z \in [0.3,0.45]$) & $-0.002 \pm 0.005$ & 0.005 & 0.001 \\      
      \hline
      High-lum.  Bin 1 ($z \in [0.15,0.3]$) & $0.004 \pm 0.005$ & 0.005 & 0.001 \\
      \hline
      High-lum. Bin 2 ($z \in [0.3,0.45]$) & $-0.009 \pm 0.005$  & 0.005 & 0.001 \\      
      \hline
    \end{tabular}
  \end{adjustbox}
  \caption{Main results for the redMaGiC photometric redshift biases, $\Delta z$, in SDSS DR8 data. Though we round statistical errors to the third decimal place, the statistical errors for the high-density sample ($\Delta z$ (stat) $\approx 0.00065$) are smaller than for high-luminosity ($\Delta z$ (stat) $\approx 0.00085$) due to more objects.}
  \label{table:sdss_final_table}
  \end{center}
\end{table}

\section{Analysis of redMaGiC Photo-z errors}
\label{sec:photozerrors}

Our main goal of this work was to measure the single parameter photometric redshift bias in each of the $dz=0.15$ redshift bins for use in DES Year 1 analyses, as shown in Section \ref{sec:results}. In this section, we go deeper into probing the full photo-z bias as a function of redshift and compare with the results of \cite{redmagicSV}, where biases are estimated for the SDSS DR8 and DES SV (science verification) redMaGiC galaxies. In \cite{redmagicSV}, for the most part only the photo-z bias of redMaGiC galaxies with spectroscopic redshifts could be analyzed, a limitation our work does not have. The purpose of this section is to identify with more precision at what redshifts the redMaGiC algorithm is biased. This study may be useful for future implementations of the redMaGiC algorithm, or in using cross-correlations for calibrating redshift distributions beyond a single shift parameter.

\begin{figure}
\begin{center}
\includegraphics[width=0.5 \textwidth]{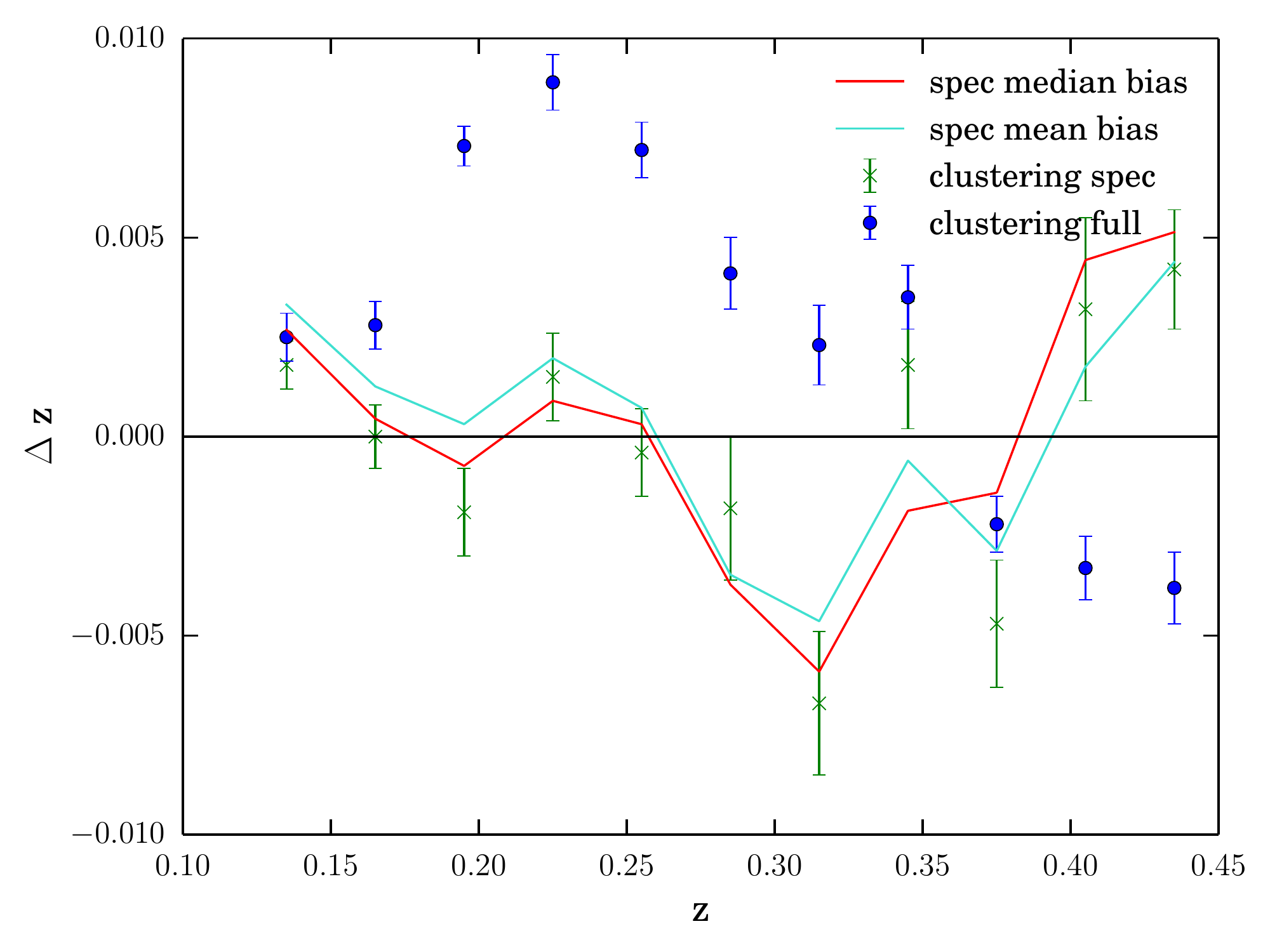}
\end{center}
\caption{Comparison of different estimates of the photometric redshift bias on SDSS redMaGiC. For the redMaGiC sample with spectroscopic redshifts, the estimates of bias by mean, median and by cross-correlations show good agreement. The cross-correlations of the full SDSS sample show similar trends of photo-z bias with redshift, but overall larger biases than on the sample with spec-z. The cross-correlations are done on $dz=0.03$ bins as selected on the redMaGiC photo-z algorithm. The mean and median redshifts shown are after cutting out all galaxies with a bias greater than 0.06. This makes the mean bias about 0.001-0.002 closer to zero, and does not affect the median. This cut is done since on these small redshift bins for the cross-correlation, such outliers will not be picked up. The comparison of estimates of means not surprisingly matches better with this cut.}
\label{fig:bias_comparison}
\end{figure}

We use our fiducial methodology for cross-correlations but this time work on thinner bins of $dz=0.03$. We first analyze SDSS redMaGiC so that we can compare directly with \cite{redmagicSV}. In Figure \ref{fig:bias_comparison}, we show the results of the cross-correlations on the $dz=0.03$ bins for the full SDSS high-density sample and the SDSS subsample with spectroscopic redshifts. The main SDSS sample has significantly larger biases at most redshifts. We also show the true photometric redshift bias of the SDSS with spec-z sample as measured by mean and median bias. Our estimates of the true median photo-z bias compares well with the results of \cite{redmagicSV} (e.g., their Figure 3). Measuring the photo-z bias by mean (which is our fiducial method in the cross-correlations) gives for the most part slightly larger amplitude bias estimates than measuring by median. The cross-correlation points for the sample with spec-z shows overall good agreement with the true mean bias and median bias, validating that our cross-correlations on these small bins are accurate. We note that we cut large photo-z outliers from the photo-z mean calculation in Figure \ref{fig:bias_comparison} to more properly compare with the cross-correlation redshift range analyzed. This changes the true mean photo-z bias by about 0.001-0.002. Based on this accuracy of the cross-correlation method, Figure \ref{fig:bias_comparison} indicates that the full SDSS redMaGiC sample indeed does have a larger photo-z bias than the sample with spec-z. The importance of this cross-correlation method to test the full redMaGiC galaxy sample rather than just the brighter galaxies that have spec-z measurements becomes clear. 

\begin{figure}
\begin{center}
\includegraphics[width=0.5 \textwidth]{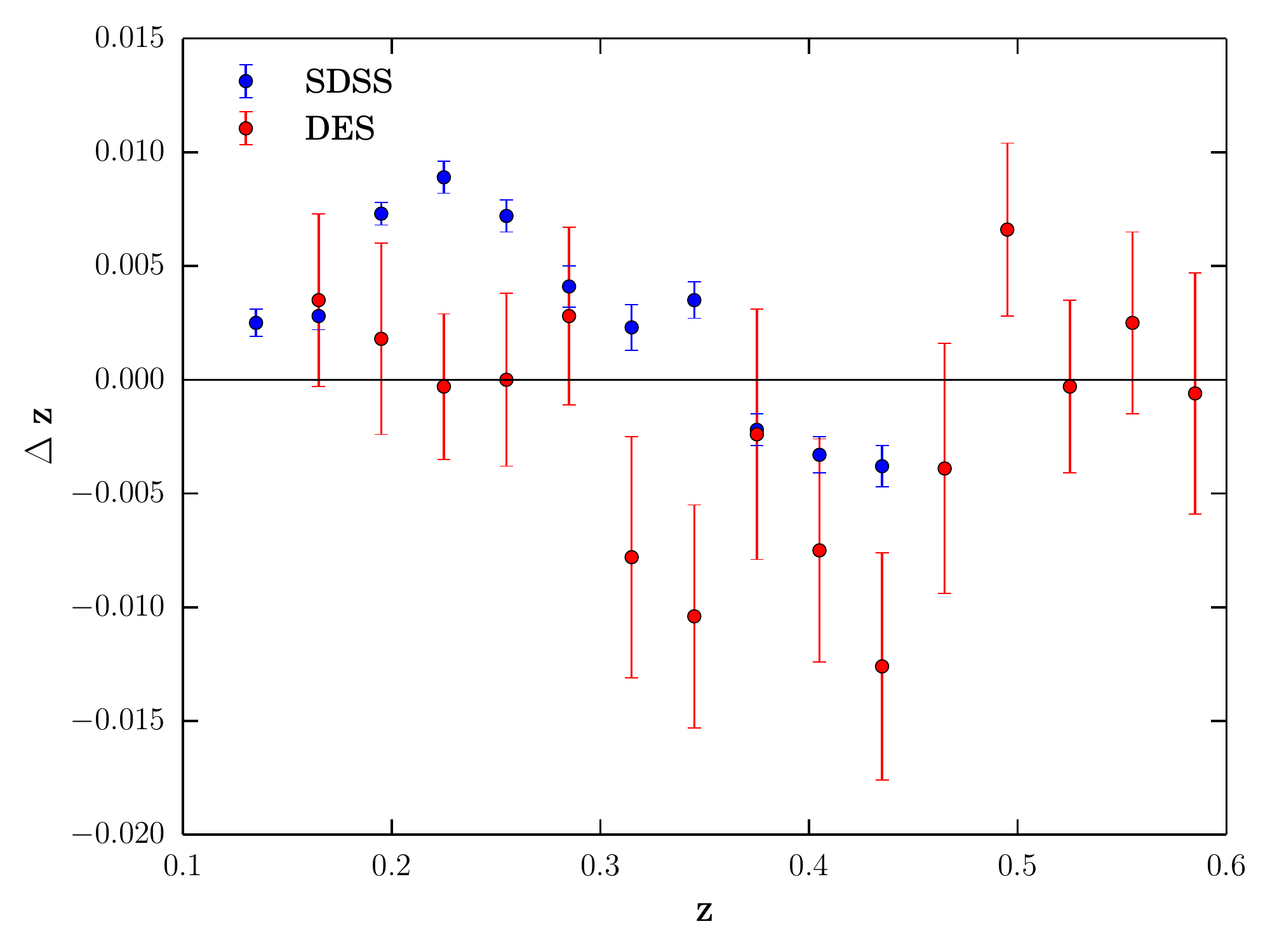}
\end{center}
\caption{Comparison of the estimated photometric redshift bias for SDSS and DES high-density samples on $dz=0.03$ size bins. The SDSS points are the same as in Figure \ref{fig:bias_comparison}.}
\label{fig:bias_comparison_des}
\end{figure}

We now compare the DES and SDSS high-density samples in the smaller $dz=0.03$ bins in Figure \ref{fig:bias_comparison_des}. As seen, the error bars in DES are significantly larger by a factor of about 5-7. Despite the larger error bars, Figure \ref{fig:bias_comparison_des} shows some interesting trends, such as a large photo-z bias at all points in Bin 2 ($z \in [0.3,0.45]$).

It is interesting to compare the apparent photo-z errors seen in Figures \ref{fig:des6}-\ref{fig:bias_comparison_des} with known issues of the redMaGiC algorithm. For example, a notable feature in the DES results is the excess of galaxies around $z=0.34$, and lack of galaxies around $z=0.42$ compared to the photo-z estimate, seen especially in the high-density results, but also to a smaller degree in the higher-luminosity results (Figure \ref{fig:des6}). This matches a photometric redshift outlier population (i.e., large photo-z bias of many individual galaxies) case mentioned in \cite{redmagicSV} (their `clump 2'). In that work, this outlier population is attributed to parallel trends in color space of dust reddening and evolution of the red sequence around $z=0.35$, which can make an e.g., $z \approx 0.3$ galaxy appear to the redMaGiC algorithm as a $z \approx 0.4$ galaxy. At other redshifts, dust reddening does not as often mimic evolution of the red sequence.

\begin{figure*}
\begin{center}
\includegraphics[width=1.0 \textwidth]{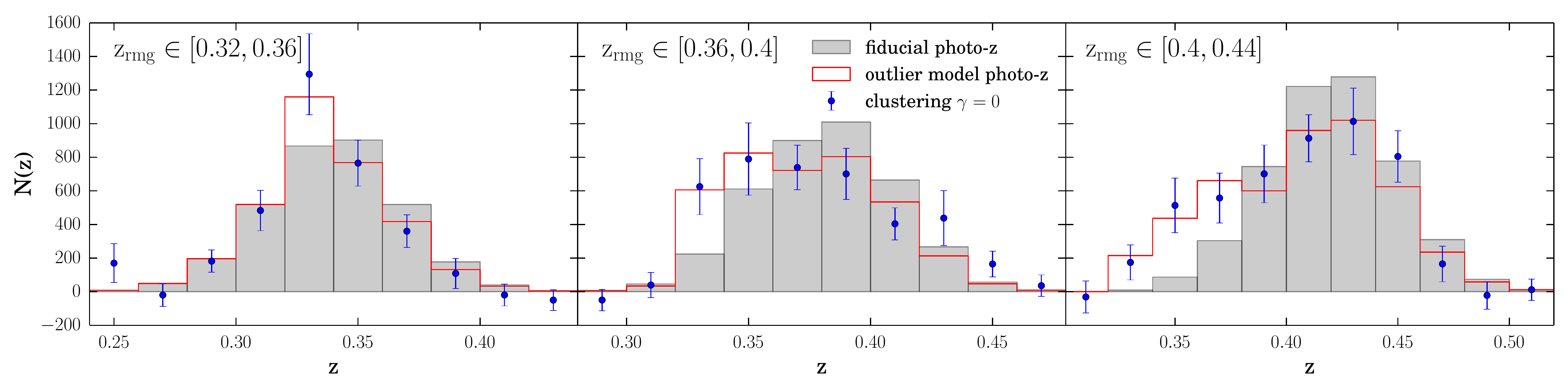}
\end{center}
\caption{Cross-correlation results for three sub-bins of our Bin 2 ($z \in [0.3,0.45]$). We create an `outlier model' to try to more accurately shift the photo-z distribution to match the clustering results, specifically trying to fit the excess of galaxies around $z=0.32-0.36$. The model is the following: randomly move $30\%$ of galaxies with $z_{\text{rmg}} \in [0.34,0.36]$ to a flat redshift distribution of $z=0.32-0.34$, similarly move $20\%$ of galaxies with $z_{\text{rmg}} \in [0.36,0.4]$ to $z=0.32-0.36$ and $20\%$ of galaxies with $z_{\text{rmg}} \in [0.4,0.44]$ to $z=0.32-0.36$. These movements are in place of the usual building of the photo-z distribution from a Gaussian sampling centered at $z_{\text{rmg}}$ with width $z_{\text{err}}$ for each galaxy.}
\label{fig:des_outlier_indiv}
\end{figure*}

\begin{figure}
\begin{center}
\includegraphics[width=0.5 \textwidth]{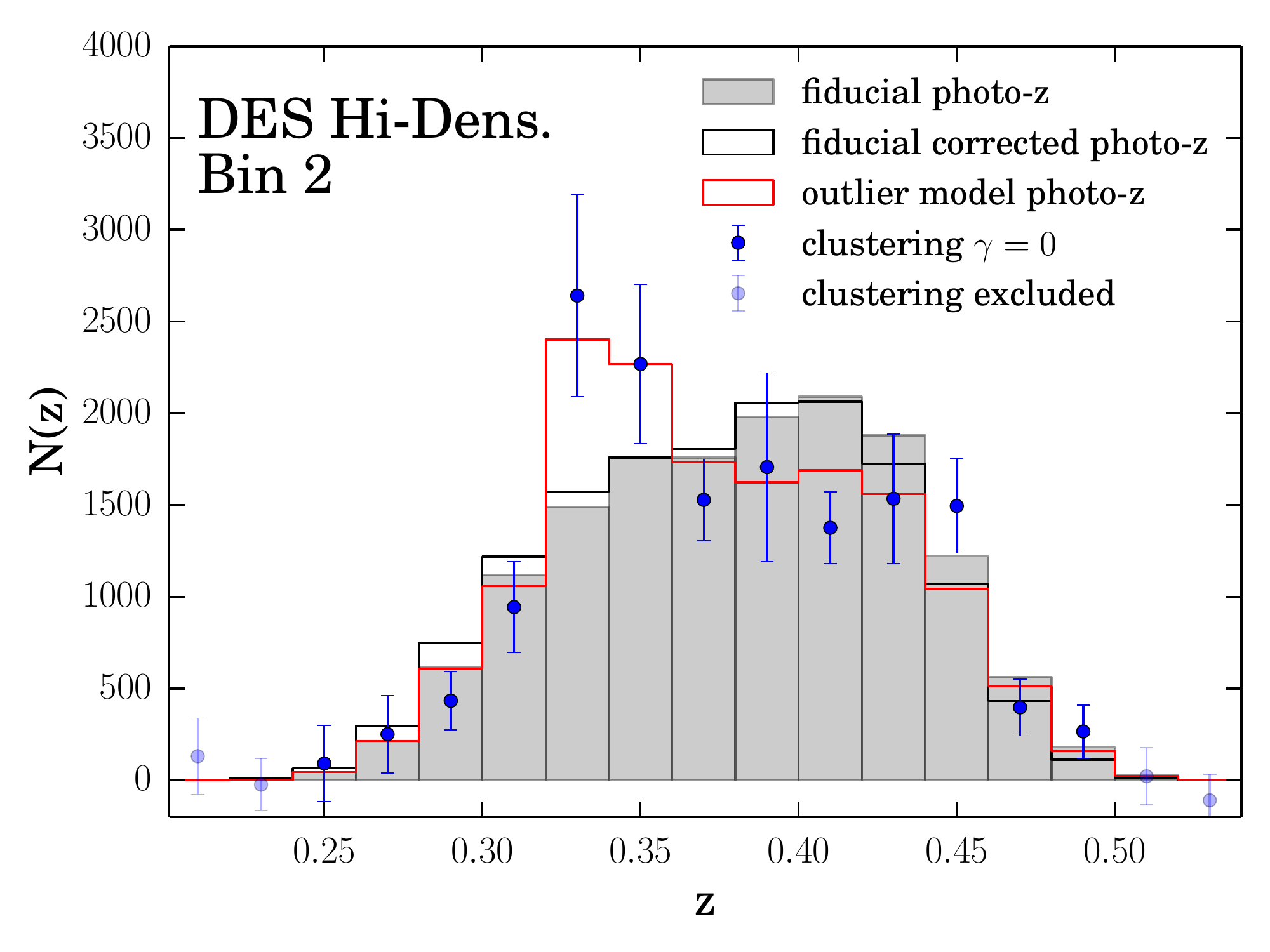}
\end{center}
\caption{Comparison of the fiducial DES Y1 high-density redMaGiC Bin 2 ($z \in [0.3,0.45]$) N(z) measurements, and shifted photo-z fit, to the outlier model described in Figure \ref{fig:des_outlier_indiv}. This improves the $\chi^2/\text{dof}$ fit to 2.2 compared to 0.7 for the fiducial corrected photo-z fit of a single shift of $\Delta z=0.005$.}
\label{fig:des_outlier_full}
\end{figure}

We can investigate where the photo-z outliers are coming from in redshift space by looking at the cross-correlations of smaller redshift bins. In Figure \ref{fig:des_outlier_indiv}, the cross-correlations of DES high-density on bins of $dz=0.04$ are shown spanning from $z=0.32-0.44$. In each plot, clear trends of excess galaxies around $z=0.33$ are present compared to the photo-z estimate. In each of the three $dz=0.04$ bins, we do a very loose matching of an `outlier model' correction to the photo-z distribution to match the clustering results, the details of which are in the caption of Figure \ref{fig:des_outlier_indiv}. Figure \ref{fig:des_outlier_full} shows the full Bin 2 ($z \in [0.3,0.45]$) with this outlier model. Across this full bin, the outlier model improves the $\chi^2/\text{dof}$ fit of the clustering data from about 2.2 to 0.7, compared to the fiducial model of applying a single shift to the photo-z distribution. These calculations use only the statistical error bars, and do not take into account the systematic uncertainties from Section \ref{sec:systematics}. The model is a rough approximation (no minimization criteria applied), but the improvement in $\chi^2$ indicates roughly these trends are present. The outlier models suggest approximately $20\%$ of galaxies in these redshift ranges are truly around $z=0.32-0.36$. This is significantly higher than e.g., the estimate of $< 5\% \ 5\sigma$ outliers (with $\sigma$ measuring the photo-z error of redMaGiC, compared with true redshift) in this redshift range, up to $z=0.45$, for DES science verification data in \cite{redmagicSV}. These types of tests indicate the ability of cross-correlations to identify specific errors in redMaGiC or other photo-z algorithms.

We can also check Figures \ref{fig:des6}-\ref{fig:bias_comparison_des} for signs of the other outlier populations (`clumps 1 and 3') in \cite{redmagicSV}. These outlier populations are attributed to undetected active star formation, meaning these galaxies are not really in the red sequence. Figures \ref{fig:sdss4} and \ref{fig:bias_comparison} indicate pretty clearly a positive photo-z bias (overprediction of low redshift) for roughly $z_{\text{pz}}=0.15-0.25$. The `clump 1' outlier population in \cite{redmagicSV}, which has photo-z biases up to +0.01, would match this trend. There are signs this trend exists in the DES data as well, though it seems to be less prevalent (Figure \ref{fig:bias_comparison_des}). There are also possibly signs in our data of `Clump 3' in \cite{redmagicSV}, which contains galaxies with $z_{\text{pz}}=0.42-0.45$ that have positive photo-z biases up to 0.015. Positive bias at this redshift range seems to be present in DES and perhaps SDSS (though only inferred from the low S/N tail for SDSS). The trends resembling clumps 1 and 3 seem much less prevalent than the one resembling `clump 2' in this work. Since our main goal was just the mean redshift of each bin, we leave more expansive modeling of redMaGiC photo-z errors to future work.

Some minor effects are worth mentioning for the analyses in this section. On these small $dz=0.03$ bins, we use a $3 \sigma_{\text{u}}$ instead of $2.5 \sigma_{\text{u}}$ cut for the SDSS samples due to the narrow distributions. This changes the results on the order of 0.001 in $\Delta z$, and generally brings the clustering estimates of the SDSS with spec-z sample closer to the true mean redshifts. For the DES samples, we specifically cut where the clustering signal is around zero or goes negative as the usual $\sigma_{\text{u}}$ cut in some bins allowed very low S/N points that seemed likely to skew the results. Variations on where to cut typically didn't affect results more than $\approx 0.002$ in $\Delta z$, though the systematics on these size bins were not as rigorously tested. It is also noteworthy that although nearly all of the broad trends in the $dz=0.15$ bins (Figures \ref{fig:des6}-\ref{fig:sdss4}) can be identified in the $dz=0.03$ bins (Figure \ref{fig:bias_comparison_des}), the overall mean biases in the main analysis are somewhat different than the inferred bias when summing the biases in the small bins even accounting for the modest change in number of galaxies across the bin. This difference is especially true in DES, where the inferred biases from summing the points in Figure \ref{fig:bias_comparison_des} differ from the fiducial results by about 0.006 in Bins 1 and 3. The summed SDSS points in Figure \ref{fig:bias_comparison_des} are within 0.002 of the fiducial results (Table \ref{table:sdss_final_table}). These differences from the fiducial analysis may be due to large redshift outliers that are found outside of a $dz=0.03$ bin and will not be picked up by the cross-correlation analysis over the $2.5-3 \sigma_{\text{u}}$ window. This will be less of an issue in the $dz=0.15$ bins, though the effect will still be present. Indeed, as mentioned in Figure \ref{fig:bias_comparison}, without cutting some outliers, the mean bias of the bin in the SDSS redMaGiC with spec-z changes in amplitude by about 0.001-0.002. The fact that the difference is larger in DES may point to statistical errors being the main factor though. Finally, we note that we again set $\gamma=0$ in the measurements in Section \ref{sec:photozerrors} for all samples. The bias evolution should be a much smaller effect across these small bins. This assumption is supported by the good match for SDSS redmagic with spec-z clustering results with the true photo-z mean biases. 

\section{Summary}
\label{sec:summary}
In this paper, we have produced constraints on the photometric redshift biases of redMaGiC galaxies using the cross-correlation method with spectroscopic galaxies as our reference sample. Our main scientific objective was to produce constraints on these biases for the redshift bins used by projects that went into the main DES Y1 cosmology paper (\cite{keypaper}, \cite{wthetapaper}, \cite{gglpaper}, \cite{xcorr}). We also looked at the SDSS redMaGiC sample which allowed us to use samples with far more galaxies (both redMaGiC and reference) to validate our methodology and to compare with the DES results. Our results on the photo-z biases of both the DES and SDSS redMaGiC samples were also used in the cosmological analysis of \cite{troughs}. We were able to study the photo-z biases as a function of redshift in more detail, providing evidence that redMaGiC biases are somewhat larger than found in \cite{redmagicSV}. We also recovered some of the known outlier trends seen in \cite{redmagicSV} in our analysis.  Overall though, the best-fit biases we found for the fiducial redshift bins are less than $\Delta z=0.01$. Our work verifies that the redMaGiC algorithm for selecting LRGs is still one of the most successful methods in creating a sample of galaxies with minimal photo-z errors. Our estimated photo-z biases for the DES high-density redMaGiC sample were small enough in magnitude to not impact the cosmological analysis of \cite{keypaper} compared to a model of zero bias with the same uncertainty as our work.  

There were two main limitations of our analysis. The first is the lack of spectroscopic galaxies to cross-correlate with across most of the DES Year 1 footprint. We only used a region about $10\%$ the size of the main sample of DES redMaGiC galaxies, overlapping the BOSS galaxies on about $124 \ \text{deg}^2$ of Stripe 82. Since there were relatively few BOSS galaxies at $z>0.6$, we could only get constraints on the first three bins used in the DES cosmological analyses. There were other smaller spectroscopic samples available in DES Y1, but these small samples would have had significantly higher statistical errors in the cross-correlations, and combining them would introduce many potential systematics compared to using a single uniformly selected sample like BOSS. Simply, to improve the analysis on DES redMaGiC, more uniformly selected spectroscopic galaxies at higher redshifts and over more of the footprint will need to be obtained.

The second limitation was our main source of systematic error, uncertainty in the galaxy bias redshift evolution of redMaGiC. This error was comparable to the statistical error of the DES redMaGiC analysis, and was approximately five times larger than the statistical error of the SDSS redMaGiC analysis. This systematic will be one that will need to continue to be addressed in future cross-correlation analyses of all types. \cite{xcorr} and \cite{xcorrtechnique} do not attempt to correct for it as the galaxy bias evolution of the weak lensing source galaxies they calibrate is more complex than redMaGiC. In this work, we do try to account for the galaxy bias evolution with some relatively conservative assumptions based on auto-correlations of redMaGiC on data and simulations. Future simulations that include accurate representations of both redMaGiC samples and spectroscopic surveys could improve calibration of systematics in this methodology, including the galaxy bias evolution.

Future datasets will lead to more opportunities of calibrating redMaGiC galaxies (and other galaxies) using cross-correlation methods. In the immediate future, the DES Year 3 dataset will cover the full footprint of DES. Compared to the year 1 analysis, this will approximately triple the area of the survey. Roughly, the overlapping area with BOSS should grow by a factor of 4-5 for year 3, allowing for a larger sample of DES redMaGiC galaxies that can be calibrated as in this work. Also now underway is the eBOSS program \citep{eboss16} which in the future could be utilized with DES for further cross-correlation. Notably, eBOSS will have more spectroscopic galaxies at higher redshifts, $z>0.6$ \citep{eboss16}. This can extend to higher redshift the calibrations possible on DES or other photometric surveys. 

Moving forward a few more years into the 2020s, there will be many potential cross-correlation applications between photometric surveys like LSST, and spectroscopic surveys like DESI \citep{desi13}, 4MOST \citep{4most}, Euclid, WFIRST and others. The cross-correlation method will continue to be a way to utilize the quantity of galaxy measurements possible with photometry and the quality redshift estimates of spectroscopic surveys. The future is also likely to bring extensions on how to use the cross-correlation method in conjunction with photo-z methods, such as identifying specific issues of photo-z algorithms as we touch on in our Section \ref{sec:photozerrors}. How best to utilize cross-correlation redshift estimates will continue to be an important area of study in cosmology.

\section*{Acknowledgements}

RC is supported by the Kavli Institute for Cosmological Physics at the University of Chicago through grant NSF PHY-1125897 and an endowment from the Kavli Foundation and its founder Fred Kavli.

Funding for the DES Projects has been provided by the U.S. Department of Energy, the U.S. National Science Foundation, the Ministry of Science and Education of Spain, the Science and Technology Facilities Council of the United Kingdom, the Higher Education Funding Council for England, the National Center for Supercomputing Applications at the University of Illinois at Urbana-Champaign, the Kavli Institute of Cosmological Physics at the University of Chicago, the Center for Cosmology and Astro-Particle Physics at the Ohio State University, the Mitchell Institute for Fundamental Physics and Astronomy at Texas A\&M University, Financiadora de Estudos e Projetos, 
Funda{\c c}{\~a}o Carlos Chagas Filho de Amparo {\`a} Pesquisa do Estado do Rio de Janeiro, Conselho Nacional de Desenvolvimento Cient{\'i}fico e Tecnol{\'o}gico and the Minist{\'e}rio da Ci{\^e}ncia, Tecnologia e Inova{\c c}{\~a}o, the Deutsche Forschungsgemeinschaft and the Collaborating Institutions in the Dark Energy Survey. 

The Collaborating Institutions are Argonne National Laboratory, the University of California at Santa Cruz, the University of Cambridge, Centro de Investigaciones Energ{\'e}ticas, Medioambientales y Tecnol{\'o}gicas-Madrid, the University of Chicago, University College London, the DES-Brazil Consortium, the University of Edinburgh, the Eidgen{\"o}ssische Technische Hochschule (ETH) Z{\"u}rich, Fermi National Accelerator Laboratory, the University of Illinois at Urbana-Champaign, the Institut de Ci{\`e}ncies de l'Espai (IEEC/CSIC), the Institut de F{\'i}sica d'Altes Energies, Lawrence Berkeley National Laboratory, the Ludwig-Maximilians Universit{\"a}t M{\"u}nchen and the associated Excellence Cluster Universe, the University of Michigan, the National Optical Astronomy Observatory, the University of Nottingham, The Ohio State University, the University of Pennsylvania, the University of Portsmouth, SLAC National Accelerator Laboratory, Stanford University, the University of Sussex, Texas A\&M University, and the OzDES Membership Consortium.

Based in part on observations at Cerro Tololo Inter-American Observatory, National Optical Astronomy Observatory, which is operated by the Association of Universities for Research in Astronomy (AURA) under a cooperative agreement with the National Science Foundation.

The DES data management system is supported by the National Science Foundation under Grant Numbers AST-1138766 and AST-1536171.
The DES participants from Spanish institutions are partially supported by MINECO under grants AYA2015-71825, ESP2015-88861, FPA2015-68048, SEV-2012-0234, SEV-2016-0597, and MDM-2015-0509, 
some of which include ERDF funds from the European Union. IFAE is partially funded by the CERCA program of the Generalitat de Catalunya.

Research leading to these results has received funding from the European Research Council under the European Union's Seventh Framework Program (FP7/2007-2013) including ERC grant agreements 240672, 291329, and 306478.
We  acknowledge support from the Australian Research Council Centre of Excellence for All-sky Astrophysics (CAASTRO), through project number CE110001020.

This manuscript has been authored by Fermi Research Alliance, LLC under Contract No. DE-AC02-07CH11359 with the U.S. Department of Energy, Office of Science, Office of High Energy Physics. The United States Government retains and the publisher, by accepting the article for publication, acknowledges that the United States Government retains a non-exclusive, paid-up, irrevocable, world-wide license to publish or reproduce the published form of this manuscript, or allow others to do so, for United States Government purposes.

\bibliographystyle{mnras}
\bibliography{xcorr,des_y1kp_short,bibliography2,bibliography3}



\section*{Affiliations}
$^{1}$ Kavli Institute for Cosmological Physics, University of Chicago, Chicago, IL 60637, USA\\
$^{2}$ Department of Astronomy and Astrophysics, University of Chicago, Chicago, IL 60637, USA\\
$^{3}$ Kavli Institute for Particle Astrophysics \& Cosmology, P. O. Box 2450, Stanford University, Stanford, CA 94305, USA\\
$^{4}$ Institut de F\'{\i}sica d'Altes Energies (IFAE), The Barcelona Institute of Science and Technology, Campus UAB, 08193 Bellaterra (Barcelona) Spain\\
$^{5}$ Jodrell Bank Center for Astrophysics, School of Physics and Astronomy, University of Manchester, Oxford Road, Manchester, M13 9PL, UK\\
$^{6}$ Department of Physics, University of Arizona, Tucson, AZ 85721, USA\\
$^{7}$ Fermi National Accelerator Laboratory, P. O. Box 500, Batavia, IL 60510, USA\\
$^{8}$ SLAC National Accelerator Laboratory, Menlo Park, CA 94025, USA\\
$^{9}$ Institute of Space Sciences, IEEC-CSIC, Campus UAB, Carrer de Can Magrans, s/n,  08193 Barcelona, Spain\\
$^{10}$ Department of Physics and Astronomy, University of Pennsylvania, Philadelphia, PA 19104, USA\\
$^{11}$ Laborat\'orio Interinstitucional de e-Astronomia - LIneA, Rua Gal. Jos\'e Cristino 77, Rio de Janeiro, RJ - 20921-400, Brazil\\
$^{12}$ Observat\'orio Nacional, Rua Gal. Jos\'e Cristino 77, Rio de Janeiro, RJ - 20921-400, Brazil\\
$^{13}$ Centro de Investigaciones Energ\'eticas, Medioambientales y Tecnol\'ogicas (CIEMAT), Madrid, Spain\\
$^{14}$ Department of Physics, Stanford University, 382 Via Pueblo Mall, Stanford, CA 94305, USA\\
$^{15}$ Institute of Astronomy, University of Cambridge, Madingley Road, Cambridge CB3 0HA, UK\\
$^{16}$ Kavli Institute for Cosmology, University of Cambridge, Madingley Road, Cambridge CB3 0HA, UK\\
$^{17}$ Universit\"ats-Sternwarte, Fakult\"at f\"ur Physik, Ludwig-Maximilians Universit\"at M\"unchen, Scheinerstr. 1, 81679 M\"unchen, Germany\\
$^{18}$ Department of Physics \& Astronomy, University College London, Gower Street, London, WC1E 6BT, UK\\
$^{19}$ Department of Physics, ETH Zurich, Wolfgang-Pauli-Strasse 16, CH-8093 Zurich, Switzerland\\
$^{20}$ Max Planck Institute for Extraterrestrial Physics, Giessenbachstrasse, 85748 Garching, Germany\\
$^{21}$ Instituci\'o Catalana de Recerca i Estudis Avan\c{c}ats, E-08010 Barcelona, Spain\\
$^{22}$ Center for Cosmology and Astro-Particle Physics, The Ohio State University, Columbus, OH 43210, USA\\
$^{23}$ Department of Physics, The Ohio State University, Columbus, OH 43210, USA\\
$^{24}$ Cerro Tololo Inter-American Observatory, National Optical Astronomy Observatory, Casilla 603, La Serena, Chile\\
$^{25}$ Department of Physics and Electronics, Rhodes University, PO Box 94, Grahamstown, 6140, South Africa\\
$^{26}$ Institute of Cosmology \& Gravitation, University of Portsmouth, Portsmouth, PO1 3FX, UK\\
$^{27}$ Instituto de Fisica Teorica UAM/CSIC, Universidad Autonoma de Madrid, 28049 Madrid, Spain\\
$^{28}$ LSST, 933 North Cherry Avenue, Tucson, AZ 85721, USA\\
$^{29}$ Observatories of the Carnegie Institution of Washington, 813 Santa Barbara St., Pasadena, CA 91101, USA\\
$^{30}$ CNRS, UMR 7095, Institut d'Astrophysique de Paris, F-75014, Paris, France\\
$^{31}$ Sorbonne Universit\'es, UPMC Univ Paris 06, UMR 7095, Institut d'Astrophysique de Paris, F-75014, Paris, France\\
$^{32}$ Department of Astronomy, University of Illinois at Urbana-Champaign, 1002 W. Green Street, Urbana, IL 61801, USA\\
$^{33}$ National Center for Supercomputing Applications, 1205 West Clark St., Urbana, IL 61801, USA\\
$^{34}$ George P. and Cynthia Woods Mitchell Institute for Fundamental Physics and Astronomy, and Department of Physics and Astronomy, Texas A\&M University, College Station, TX 77843,  USA\\
$^{35}$ Department of Physics, IIT Hyderabad, Kandi, Telangana 502285, India\\
$^{36}$ Department of Astronomy/Steward Observatory, 933 North Cherry Avenue, Tucson, AZ 85721-0065, USA\\
$^{37}$ Jet Propulsion Laboratory, California Institute of Technology, 4800 Oak Grove Dr., Pasadena, CA 91109, USA\\
$^{38}$ Department of Astronomy, University of Michigan, Ann Arbor, MI 48109, USA\\
$^{39}$ Department of Physics, University of Michigan, Ann Arbor, MI 48109, USA\\
$^{40}$ Institute of Space Sciences (ICE, CSIC) \& Institut d'Estudis Espacials de Catalunya (IEEC), Campus UAB, Carrer de Can Magrans, s/n,  08193 Barcelona, Spain\\
$^{41}$ Santa Cruz Institute for Particle Physics, Santa Cruz, CA 95064, USA\\
$^{42}$ Astronomy Department, University of Washington, Box 351580, Seattle, WA 98195, USA\\
$^{43}$ Australian Astronomical Observatory, North Ryde, NSW 2113, Australia\\
$^{44}$ Argonne National Laboratory, 9700 South Cass Avenue, Lemont, IL 60439, USA\\
$^{45}$ Departamento de F\'isica Matem\'atica, Instituto de F\'isica, Universidade de S\~ao Paulo, CP 66318, S\~ao Paulo, SP, 05314-970, Brazil\\
$^{46}$ Department of Astronomy, The Ohio State University, Columbus, OH 43210, USA\\
$^{47}$ Brookhaven National Laboratory, Bldg 510, Upton, NY 11973, USA\\
$^{48}$ School of Physics and Astronomy, University of Southampton,  Southampton, SO17 1BJ, UK\\
$^{49}$ Instituto de F\'isica Gleb Wataghin, Universidade Estadual de Campinas, 13083-859, Campinas, SP, Brazil\\
$^{50}$ Computer Science and Mathematics Division, Oak Ridge National Laboratory, Oak Ridge, TN 37831\\

\bsp	
\label{lastpage}
\end{document}